\documentclass[preprint,12pt,authoryear]{elsarticle}%%单列显示
\usepackage[a4paper, top=2.5cm, bottom=2.5cm, left=2cm, right=2cm]{geometry} % 设置页边距,注意若需要使用双列排版时，需要注释这句话，不然要报错

\usepackage{amssymb}
\usepackage{amsmath}
\usepackage{multirow}%
\usepackage{amsmath,amssymb,amsfonts}%
\usepackage{amsthm}%
\usepackage{mathrsfs}%
\usepackage[title]{appendix}%
\usepackage{xcolor}%
\usepackage{textcomp}%
\usepackage{manyfoot}%
\usepackage{booktabs}%
\usepackage{algorithm}%
\usepackage{algorithmicx}%
\usepackage{algpseudocode}%
\usepackage{listings}%
\usepackage{natbib}
\usepackage{booktabs}
\usepackage{bm}
\usepackage{hyperref}
\usepackage{subcaption}
\usepackage{booktabs} % 为美观的表格
\usepackage{tabularx} % 支持自适应宽度的表格
\usepackage{caption}  % 支持更好的表格和图形标题

\def\be{\begin{equation}}
	\def\ee{\end{equation}}
\def\bea{\begin{eqnarray}}
	\def\eea{\end{eqnarray}}

\journal{High Energy Astrophysics}

\graphicspath{{Graphics/}}

\def\be{\begin{equation}}
	\def\ee{\end{equation}}
\def\bea{\begin{eqnarray}}
	\def\eea{\end{eqnarray}}

\begin{document}

\begin{frontmatter}

%% Title, authors and addresses

%% use the tnoteref command within \title for footnotes;
%% use the tnotetext command for theassociated footnote;
%% use the fnref command within \author or \affiliation for footnotes;
%% use the fntext command for theassociated footnote;
%% use the corref command within \author for corresponding author footnotes;
%% use the cortext command for theassociated footnote;
%% use the ead command for the email address,
%% and the form \ead[url] for the home page:
%% \title{Title\tnoteref{label1}}
%% \tnotetext[label1]{}
%% \author{Name\corref{cor1}\fnref{label2}}
%% \ead{email address}
%% \ead[url]{home page}
%% \fntext[label2]{}
%% \cortext[cor1]{}
%% \affiliation{organization={},
%%            addressline={},
%%            city={},
%%            postcode={},
%%            state={},
%%            country={}}
%% \fntext[label3]{}

\title{Gamma-Ray Bursts Calibrated from the Observational $H(z)$ Data in Artificial Neural Network Framework}

%% use optional labels to link authors explicitly to addresses:
%% \author[label1,label2]{}
%% \affiliation[label1]{organization={},
%%             addressline={},
%%             city={},
%%             postcode={},
%%             state={},
%%             country={}}
%%
%% \affiliation[label2]{organization={},
%%             addressline={},
%%             city={},
%%             postcode={},
%%             state={},
%%             country={}}

\author[label1,label3]{Zhen Huang\fnref{email1}}
\author[label1]{Zhiguo Xiong}
\author[label1]{Xin Luo}
\author[label1]{Guangzhen Wang}
\author[label2]{Yu Liu\corref{cor2}}
\author[label1]{Nan Liang\corref{cor1}}
%% Author affiliation
\affiliation[label1]{organization={Guizhou Key Laboratory of Advanced Computing},%Department and Organization
            addressline={Guizhou Normal University},
            city={Guiyang},
            postcode={550025},
            state={Guizhou},
            country={China}}
\affiliation[label2]{organization={School of Physical Science and Technology},
	        addressline={Southwest Jiaotong University},
	        city={Chengdu},
	        postcode={611756},
         	state={Sichuan},
        	country={China}}
\affiliation[label3]{organization={School of Cyber Science and Technology},%Department and Organization
	        addressline={Guizhou Normal University},
	        city={Guiyang},
          	postcode={550025},
        	state={Guizhou},
	        country={China}}

\cortext[cor1]{Corresponding author. Email: liangn@bnu.edu.cn}
\cortext[cor2]{Corresponding author. Email: lyu@swjtu.edu.cn}
%\cortext[cor3]{First author. Email: zhen\_huang@gznu.edu.cn}
%\cortext[cor1]{Corresponding author.}
%\cortext[cor2]{Corresponding author.}
\fntext[email1]{Email: zhen\_huang@gznu.edu.cn}
%\fntext[email2]{Email: zhiguoxiong@gznu.edu.cn}
%\fntext[email3]{Email: xin\_luo@gznu.edu.cn}
%\fntext[email4]{Email: guangzhenwang@gznu.edu.cn}
%\fntext[email5]{lyu@swjtu.edu.cn}
%\fntext[email6]{liangn@bnu.edu.cn}

%% Abstract
\begin{abstract}
In this paper, we calibrate the luminosity relation of gamma-ray bursts (GRBs) from an Artificial Neural Network (ANN) framework for reconstructing the Hubble parameter \unboldmath{$H(z)$} from the latest observational Hubble data (OHD) obtained with the cosmic chronometers method in a cosmology-independent way. We consider the physical relationships between the data to introduce the covariance matrix and KL divergence of the data into  the loss function and calibrate the Amati relation ($E_{\rm p}$--$E_{\rm iso}$) by selecting the optimal ANN model with the A219 sample and the J220 sample  at low redshift. Combining the Pantheon+  type Ia supernovae (SNe Ia) sample and Baryon acoustic oscillations (BAOs) from Dark Energy Spectroscopy Instrument (DESI) with  GRBs at  high redshift  in the Hubble diagram by Markov Chain Monte Carlo numerical method, we find that the $\Lambda$CDM model is preferred over the $w$CDM and CPL models with joint constraints by the Akaike Information Criterion and Bayesian Information Criterion.
\end{abstract}

%%%Graphical abstract
%\begin{graphicalabstract}
%%\includegraphics{grabs}
%\end{graphicalabstract}
%
%%%Research highlights
%\begin{highlights}
%\item Research highlight 1
%\item Research highlight 2
%\end{highlights}

%% Keywords

\end{frontmatter}

\section{Introduction}
Gamma-ray bursts (GRBs) are the most powerful high-energy phenomena in the universe, %releasing intense gamma rays over extremely short timescales, essential
which can be used to probe the universe up to \(z \sim 9\) \citep{Salvaterra2009,Cucchiara2011}, far exceeding the maximum redshift of Type Ia supernovae (SNe Ia) at \(z \sim 2\) \citep{Scolnic2018,Scolnic2022}. %which makes GRBs particularly valuable for probing the universe at higher redshifts.
The luminosity relations of GRBs %, which link observable properties of gamma-ray emissions to their intrinsic luminosity or energy
\citep{Amati2002,Ghirlanda2004a,Yonetoku2004,Liang2005,Dainotti2008,Dainotti2016,Izzo2015} have enabled their use as cosmological tools \citep{Dai2004,Ghirlanda2004b,Liang2006,Ghirlanda2006,Schaefer2007,Wang2015,Dainotti2018,Luongo2021a,Han2024,Li2024,Bargiacchi2025} and \cite{Hu2021,Wang2022,Cao2022a,Cao2022b}. %to investigate the nature of DE (dark energy).
In early studies, the luminosity relations usually calibrated  by assuming a specific cosmological model due to the absence of a low-redshift GRB sample \citep{Dai2004}, which can introduce the circularity problem. %to constrain cosmological models.
\cite{Liang2008} proposed a model-independent calibration method by interpolating GRB data from low-redshift SNe Ia observations; therefore GRBs can be used to constrain cosmological parameters without assuming a prior cosmological model \citep{Capozziello2008,Capozziello2009,Wei2009,Wei2010,Liang2010,Liang2011,Wang2016,Liu2022b}.

Another strategy to mitigate the circularity problem is the simultaneous fitting method \citep{Amati2008,Wang2008}, where the parameters of the GRB relations and the cosmological model are determined jointly.  %Although this method requires a predefined cosmological model, r
Recent findings suggest that the GRB relation parameters remain consistent across different cosmological models, implying that GRB data can be standardized within error margins \citep{Khadka2020}.
GRB luminosity relations can also be calibrated using other observational datasets. For instance, \cite{Amati2019} employed the observational Hubble data (OHD) derived from the cosmic chronometers (CC) method and used a B\'{e}zier parametric curve to calibrate the GRB \(E_{\rm p}\)-\(E_{\rm iso}\) correlation (commonly known as the Amati relation) \citep{Amati2002} %This approach has led to the construction of a Hubble diagram for 193 GRBs and has been used in various studies to
and constrain cosmological models \citep{Montiel2021,Luongo2021b,Luongo2023,Muccino2023}.

The calibration of GRBs using cosmological data can be approached in various reconstruction techniques: such as the interpolation technique \citep{Liang2008,Liu2022b}, polynomial fitting \citep{Kodama2008}, iterative procedures \cite{LiangZhang2008}, local regression techniques \citep{Cardone2009,Demianski2017,Demianski2021}, cosmographic approaches \citep{Capozziello2010,Gao2012,Dai2021}, two-step methods minimizing the reliance on SNe Ia \citep{Izzo2015, Muccino2021}, the Pad\'{e} approximation \citep{Liu2015}, and the B\'{e}zier parametric approach \citep{Amati2019}.
At the other hand, non-parametric methods, demonstrating superior accuracy in reducing the errors typically associated with the aforementioned techniques have been introduced to reconstruct from observational data. %Notably, \cite{Postnikov2014} explored the evolution of the cosmological equation of state through non-parametric reconstruction using high-redshift GRBs.
One of the most promising non-parametric approaches is the Gaussian Process (GP), which is a data smoothing technique grounded in Bayesian statistics. In recent years, the GP method has gained significant traction as a model-independent regression tool in cosmology, e.g., \cite{Seikel2012a,Seikel2012b,Li2018,Pan2020,Li2021,Sun2021,Mu2023a,Mu2023b} %Seikel2013,Busti2014,,
and \cite{Zhang2024a}.
However, GP faces several challenges to reconstruct $H_0$ from observational data \citep{Wei2017}.
Moreover, not all cosmological datasets follow a normal distribution, which can effect the reliability of reconstructed $H(z)$ data \citep{Zhang2024b}. %Furthermore, \cite{Wei2017} observed that the GP method is highly sensitive to the choice of the fiducial Hubble constant, $H_0$, with significant deviations in the results depending on its value. \cite{Zhou2019} also cautioned against using GP for the reconstruction of observational Hubble data (OHD) and SNe Ia, as the results may be influenced by the selection of kernel functions.

Recently, it should be note that the potential use of machine learning (ML) algorithms for cosmological use with GRBs.
ML are a set of technologies that learn to make predictions and decisions by training with a large amount of the observational data, which are a collection of processing units designed to identify underlying relationships in input data. %therefore, when an appropriate network is chosen, the model created using ML can accurately depict the distribution of the input data in a completely data-driven way.
The ML methods have shown outstanding performance in solving cosmological problems in both accuracy and efficiency to provide powerful tools and methods for cosmological research \citep{Luongo2021b}. \cite{Zhang2025} used the Pantheon+ sample \citep{Scolnic2022} to calibrate the Amati relation %from the A118 GRB sample
with  ML  approaches. % including 49 GRBs from Fermi catalog.
Moreover, the concept of Artificial neural networks (ANN), %was first proposed by \cite{McCulloch1943}. The ANN method,
completely driven by data, allows us to reconstruct a function from any kind of data without assuming a parametrization of the function.
\cite{WangGJ2020} proposed an ANN  approach  for reconstructing functions from observational data with Hubble parameter measurements $H(z)$ and the distance-redshift relation of SNe Ia.
%\cite{Wang2020} proposed a new nonparametric approach for reconstructing a function from observational data using an Artificial Neural Network (ANN).
\cite{Escamilla-Rivera2020} used the Recurrent Neural Networks (RNN) and the Bayesian Neural Networks (BNN) to reduce the computation load of expensive codes for dark energy models; which have subsequently been used to calibrate the GRB relations \citep{Escamilla-Rivera2022,Tang2021}.
%\cite{Qi2020} utilized GP and ANN methods to achieve the distance construction from $H(z)$ data and then measure the cosmic curvature parameter $\Omega_k$ in a model-independent way.
\cite{Dialektopoulos2022} noted that another advantage of the ANN method is %that its $H(z)$ reconstruction is largely independent of priors, while GP is significantly affected by different priors.
%These fundamental distinctions arise because
the ANN structure makes fewer assumptions than GP in this context, providing a more authentic representation of cosmological parameters.
\citet{Gomez-Vargas2023} reconstructed cosmological functions using ANNs based on observational measurements with minimal theoretical and statistical assumptions %, and tested the proposed method
with data from CC, $f\sigma_8$ measurements, and %the distance modulus of
SNe Ia. \cite{Zhang2024b} found that %while training data and the objective function guide the learning process,
ANNs do not inherently rely on prior beliefs about the data distribution.

Recently,
\cite{Khadka2021} compiled  the A220 sample in which the A118 sample with the smallest intrinsic dispersion.
\cite{Liang2022}  calibrated the Amati relation with the  A219 sample updated from the A220 sample.
\cite{Wang2024} constrained the emergent dark energy models with the A118 sample, OHD and baryon acoustic oscillation (BAO) data at intermediate redshift.
\cite{NL2024} used the A118 GRB sample and  the Pantheon+ sample to test the phenomenological interacting dark energy model.
More recently,
\cite{Jia2022} tested the Amati relation by using 221 GRBs (the J221 sample) based on the previous data including 49 GRBs from \emph{Fermi} catalog.
\cite{Xie2023} used a GP  approach  to calibrate the Amati relation with the J221 sample from the Pantheon+ sample. % including 49 GRBs from Fermi catalog.
\cite{Cao2024} used  the updated J220 GRB sample from \cite{Jia2022} to  simultaneously constrain Amati correlation parameters and cosmological parameters from a joint analysis of OHD and BAO data.
%\cite{WL2024} presented a sample of long GRBs  from 15 years of the Fermi-GBM catalogue with redshift measured at $0.0785\le z\le8.2$.
\cite{Shah2024}  used a novel deep learning framework called LADDER (Learning Algorithm for Deep Distance Estimation and Reconstruction) to reconstruct the cosmic distance ladder  with the A219 sample.

In this paper, we propose an ANN framework  to  calibrate the Amati relation  with the A219 sample \citep{Liang2022,Khadka2021} and the J220 sample \citep{Cao2024,Jia2022}  from the latest OHD at low redshift. Combining GRB data at high redshift with the Pantheon+ sample and  BAOs, we constrain dark energy models in a flat space by the MCMC method.	%Finally, we also compare the results of GP method and ANN method.

\section{Reconstruction From ANN}
\begin{figure*}
	\centering
	\includegraphics[width=360pt]{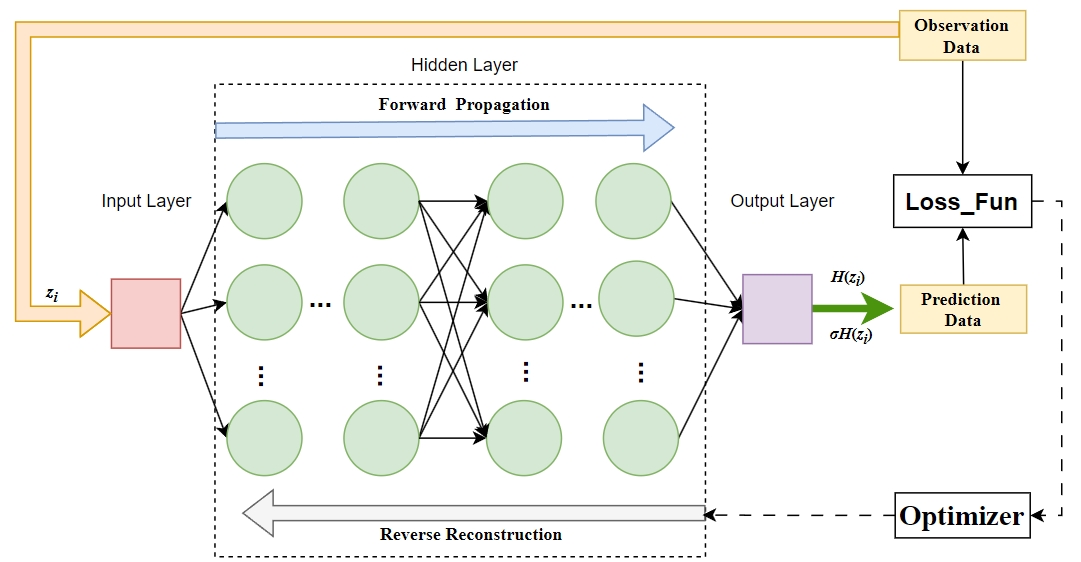}
	\caption{The structure of the ANN used in this work. The input is the redshift $z$ of a Hubble parameter $H(z)$, and the outputs are the predicted $H(z)$  and $\sigma_{H(z)}$ values. %In FFNN, each neuron is divided into different groups according to the order of receiving information, and each group can be regarded as a nerve layer.
			% The information in the whole network is propagated in one direction, %and there is no reverse information propagation.while the arrow in the graph represents error backpropagation with the loss function. %We can think of the neural network as a function, which can realize the complex mapping from the input space to the output space through multiple compositions of simple nonlinear functions.
		The thick arrows in the hidden layers represent the forward propagation of the data and the white arrows represent the reverse reconstruction of the error with the loss function.
	}
	\label{fig_ANN_structure}
\end{figure*}

%\subsection{Artificial neural networks}

%The general structure of an ANN is composed of an input layer that connects to a hidden layer or a series of successive hidden layers, and an output layer, where the elements of each layer are known as neurons. The input layer consists of the features of the dataset, associating the corresponding values as an input signal for the neurons of the input layer. On the other hand, the output layer consists of neurons whose values must be evaluated by an error function (also loss function) that measures the difference between the values predicted by the ANN and the actual values of the dataset. We illustrate such a structure for a one-hidden-layer ANN with n neurons in
%Fig. \ref{fig_ANN_structure}. In this work, for each redshift point at the input layer, the ANN outputs a predicted Hubbble parameter $H(z)$ and its corresponding uncertainty $\sigma_{H(z)}$.
%ANN is a computational model inspired by biological neural networks designed to mimic the way the human brain processes information.
%At present, the three common neural network structures are feedforward neural network, feedback neural network (FFNN) and graph network.
In this paper, we consider the feedforward neural network (FFNN) with a structure consisting of input and output layers connected to a hidden layer or a series of successive hidden layers, where elements of each layer are called neurons.
%We illustrate such a structure for a double-hidden-layer ANN with n neurons in Fig. \ref{fig_ANN_structure}.
We show the structure of a neural network with $L$-layers in Fig. \ref{fig_ANN_structure}.
The input layer consists of the features of the dataset, with each neuron representing a feature or variable.
The output layer is mainly responsible for the output of the final result, the specific form of which depends on the task,% (classification, regression, etc.),
whose value must be evaluated by the error function, also known as the loss function, which measures the difference between the predicted value of ANN and the actual value of the data set.

The neurons in the hidden layers for each layer receive the output of the neurons in the previous layer and output it to the neurons in the next layer.
%One of the main roles is
The hidden layer in ANN is responsible for neural network learning and data processing, which can have one or more layers.
For each point in the input layer, the hidden layer performs a linear transformation (consisting of linear weights and biases) and nonlinear activation on it, and then passes the inferred results to the output layer. %, a process that we call forward propagation.
 For the forward propagation in the hidden layers with $L$-layers, the output ${a}^{(l)}$ at layer $l$ is % , $L$ indicates the output layer.
\begin{equation}
	\mathbf{\emph{a}}^{(l)}=f^{(l)}\left(\mathbf{\emph{W}}^{(l)} \mathbf{a}^{(l-1)}+\mathbf{\emph{b}}^{(l)}\right), \quad l=1,2, \ldots, L
\end{equation}
here, $f(x)$ is the activation function.\footnote{In regression problems, the output layer uses the identity function, while hidden layers use nonlinear activation functions. We use the Exponential Linear Unit (ELU) \citep{Clevert2015},
	%\begin{eqnarray}\label{ELU}
	$f(x)=
	\begin{cases}
		x, & x > 0 \\
		\alpha(e^x-1), & x \leq 0 \\
	\end{cases}, $
	%\end{eqnarray}
	where $\alpha$ is a positive hyperparameter controlling the saturation for negative inputs, which are set to 1 for convenience.} $W$ and $b$ represent the weights and the biases.
In order to better minimize the difference between the predicted value $\hat{Y}$ and the true value $Y$, we calculate a loss function ($\mathcal{L}$) to measure the difference between the predicted value and the true value.
When getting the value of the loss function, we need to pass the parameter and gradient information to an optimizer\footnote{There are four common optimizers: SGD, Momentum, AdaGrad, Adam. In our work, we adopt Adam \citep{Kingma2014} as the optimizer, which can accelerate the update of parameters. %Adam realizes efficient search of parameter space by combining the advantages of Momentum with AdaGrad.
However, it should be noted that the results will vary depending on the problem to be solved.%, in fact, no one method performs well on every issue.
}
which is responsible for the update of the network parameters in each iteration.
For the reverse reconstruction in in the hidden layers, the weight and bias is updated by
\begin{equation}
	W^{(l)} \leftarrow W^{(l)} - \eta \nabla W^{(l)} \mathcal{L}, \quad b^{(l)} \leftarrow b^{(l)} - \eta \nabla b^{(l)} \mathcal{L},
\end{equation}
here $\eta$ refers to the learning rate.

ANNs can handle a wider range of data types, but still require careful design of the network structure and loss function. %In setting up the loss function,
It should be noted that different loss functions will also affect the final prediction results.
\cite{WangGJ2020} used the Mean Absolute Error (MAE)\footnote{MAE without accounting for the data uncertainties is defined as: $
	% \rm{MSE}=\frac{1}{\emph{N}}\sum^{N}_{\emph{\emph{i}}=1}(\emph{Y}_\emph{i}-\hat{\emph{Y}_\emph{i}})^2,
	\textrm{MAE} = \frac{1}{N} \sum_{i=1}^{N} (H_{\rm obs}(z_i)-H_{\rm pred}(z_i)).
	$} as a loss function.
\cite{Gomez-Vargas2023} used the values of the mean squared error (MSE)\footnote{MSE without accounting for the data uncertainties is defined as: $
	% \rm{MSE}=\frac{1}{\emph{N}}\sum^{N}_{\emph{\emph{i}}=1}(\emph{Y}_\emph{i}-\hat{\emph{Y}_\emph{i}})^2,
	\textrm{MSE} = \frac{1}{N} \sum_{i=1}^{N}(H_{\rm obs}(z_i)-H_{\rm pred}(z_i))^2.
	$}.
%When dealing with two or more variables,
ANN will organize multiple target outputs into a tensor, and then sum (or average) their loss function values. %but  MSE is actually used as a loss function in its code.
However, it is not possible to simply add the two loss values when finding the loss function.\footnote{$\sigma_{H(z)}$ usually represents the uncertainty (standard deviation) of $H(z)$ and is used to construct a Gaussian distribution to describe the statistical properties of the measured value. For example,  measurement of $H(z)$ value is generally assume that obey gaussian distribution: $ H_{\text{obs}}(z) \sim \mathcal{N}(H(z), \sigma_{H(z)}^2)$. %where $ H_ {true}(z)$ is the true value.
	In this way, the Gaussian distribution is used to describe the statistical error of the measured value, rather than $H(z)$ itself.}
%Fig. \ref{fig_Re_Wang2020} is the $H(z)$ reconstructed by ANN using the latest 33 OHD data with reference to MSE.
%\begin{figure}
%	\centering
%	\includegraphics[width=230pt]{./img/Re_Wang2020}
%	\caption{
	%		The reconstructed results from OHD through ANN. The black curves present the reconstructed function with the 1$\sigma$ uncertainty.
	%		The dashed green curve is obtained with $H_0$ = 67.36 km $s^{-1}$ ${{\rm Mpc}}^{-1}$, $\Omega_m$ = 0.315 \citep{Plank2020}, and the solid green curve is obtained with $H_0$ = 74.3 km $s^{-1}$ ${{\rm Mpc}}^{-1}$, $\Omega_m$ = 0.298 \citep{Scolnic2018}.}
%	\label{fig_Re_Wang2020}
%\end{figure}
Recently, \cite{Chen2024} used two sets of neural networks to train  $H(z)$ and their uncertainty individually.
\cite{Shah2024} used KL divergence as a loss function to consider the physical meaning represented by $H(z)$ and $\sigma_{H(z)}$.
KL divergence can maintain consistency between the predicted and observed distributions, also known as relative entropy, is a method to describe the difference between two probability distributions, which defined as:
\begin{equation}
	D_{\rm KL}(P \| Q)=\sum_{i} P(i) \log \left(\frac{P(i)}{Q(i)}\right)
\end{equation}
where $P$ and $Q$ are the true distribution  and the fitting distribution.
%Where $P$ represents the true distribution and $Q$ represents the fitting distribution of $P$. %Fig. \ref{fig_Re_LADDER2024} shows the reconstructed $H(z)$ after replacing the loss function with KL divergence.
%\begin{figure}
%	\centering
%	\includegraphics[width=230pt]{./img/Re_LADDER2024}
%	\caption{
	%		The reconstructed results from OHD through ANN. The blue curves present the reconstructed function with the 1$\sigma$ uncertainty.
	%		The dashed green curve is obtained with $H_0$ = 67.36 km $s^{-1}$ ${{\rm Mpc}}^{-1}$, $\Omega_m$ = 0.315 \citep{Plank2020}, and the solid green curve is obtained with $H_0$ = 74.3 km $s^{-1}$ ${{\rm Mpc}}^{-1}$, $\Omega_m$ = 0.298 \citep{Scolnic2018}.}
%	\label{fig_Re_LADDER2024}
%\end{figure}

%\subsection{The $H(z)$ Data} %Optimal ANN Model with The Mock Data}
In this work, we consider the physical relationships between the data in order to introduce the covariance matrix and KL divergence of the data.
This approach not only allows for model-independent reconstruction but also enhances the analysis of cosmological models by considering the relationships between data, thereby improving the interpretation of their behavior.
For the $H(z)$ data, we use the latest OHD in the range including the recent 31 point at $0.179\leq z\leq1.965$ \citep{Jimenez2003,Simon2005,Stern2010,Moresco2012,Moresco2015,Moresco2016,Zhang2014,Ratsimbazafy2017}, and the new point at $z=0.80$ \citep{Jiao2022} \footnote{More recently, \cite{Borghi2022} explored a new approach to obtain a piont OHD at $z=0.75$.
\cite{Jiao2022} proposed a similar approach to obtain a point at $z=0.80$. For these two measurements are not fully independent and their covariance is not clear, \cite{Zhang2022} only use the point \cite{Jiao2022} %,  which taking advantage of the $~1/\sqrt 2$ fraction of systematic uncertainty,
with other 31 OHD to calibrate of HII Galaxies.} and the point %$H(z)=135\pm65$
at $z=1.26$ \citep{Tomasetti2023}. We consider the total covariance matrix, which combines the statistical and systematic errors \citep{Moresco2020} of 15 $H(z)$ estimates in the range $0.179 < z < 1.965$ \citep{Moresco2012, Moresco2016, Moresco2015}. %which are summarized in Table \ref{OHD}.
At first, we considered more complexity in the observed data sets by defining a new loss function analogous to the MCMC log-likelihood for the covariance matrix of data \citep{Dialektopoulos2023}, which allows the model to properly account for the data uncertainties.
\begin{equation}
	% \rm{MSE}=\frac{1}{\emph{N}}\sum^{N}_{\emph{\emph{i}}=1}(\emph{Y}_\emph{i}-\hat{\emph{Y}_\emph{i}})^2,
	%	L_{\chi^2} = \sum_{i=1}^{N} (H_{\rm obs}(z_i)-H_{\rm pred}(z_i))C_{ij}(H_{\rm obs}(z_i)-H_{\rm pred}(z_i)).
	\mathrm{\emph{L}}_{\chi^{2}}=\sum_{i;j}^N\left[H_{\mathrm{obs}}\left(z_{i}\right)-H_{\mathrm{pred}}\left(z_{i}\right)\right]^{\mathrm{T}} \mathrm{\emph{C}}_{i j}^{-1}\left[H_{\mathrm{obs}}\left(z_{j}\right)-H_{\mathrm{pred}}\left(z_{j}\right)\right].
\end{equation}
Where ${C}_{ij}$ is the total covariance matrix of 15 $H(z)$\footnote{\url{https://gitlab.com/mmoresco/CCcovariance/-/tree/master?ref_type=heads}}.
To further ensure that the model provides high confidence in the probability distribution of the outputs while directly optimizing the deviation between the predicted and true values, %to ensure the accuracy of the output,
we adopted KL divergence combined with the MCMC log-likelihood  as the loss function for optimization,
\begin{equation}
	\mathcal{L_{\rm com}}=D_{\rm KL}+L_{\chi^2}.
\end{equation}

The degree of complexity of the ANN should reflect the structure of the physical process which is producing the data.
In order to obtain optimal ANN model before training with the actual data set, we generate a mock data set\footnote{Following the method proposed by \cite{MaZhang2011}, the simulated data can be given by assuming the spatially-flat $\Lambda$CDM model
%\begin{equation}
%	H(z)=H_{0}\sqrt{\Omega_{\rm m}(1+z)^3 + 1 - \Omega_{\rm m}}
%\end{equation}
%Here we use $H_{0} = 70 km/s/Mpc$ and $\Omega_{\rm m} = 0.3$.
%Note that the redshift distribution histogram of the actual OHD shown in the top pannel of Fig. \ref{redshift_distribution}. Therefore,
with the simulated redshift distribution subjects to a Gamma distribution %shown in the bottom of Fig. \ref{redshift_distribution}. The latest 32 OHD obtained with the CC method is summarized in Table \ref{OHD}.
for the redshift distribution histogram of the actual OHD.
Assuming a linear model for the error of $H(z)$ to fit a first degree polynomial function of redshift,
%The mean fitting function is found to be $\sigma^{0}_H (z)=15+9.82z$, while the symmetric upper and lower error bands are respectively specified by $\sigma^{+}_H (z)=25+16.82z$ and $\sigma^{-}_H (z)=5+2.82z$.
%These fitting functions are also depicted in the Fig. \ref{sigma_Hz}, in which one could easily observe that the majority of the data points are included in the area enclosed by the $\sigma^{+}_H (z)$ and $\sigma^{-}_H (z)$ functions (blue solid lines).
%Now, we randomly generate the error for our $H(z)$ mock data points, in which
and %the error $\tilde{\sigma}_H(z)$ follows
the Guassian distribution of the error $\tilde{\sigma}_H(z)$: $\mathcal{N}(\sigma^{0}_H (z), \varepsilon_H(z))$, where $\varepsilon_H(z)=(\sigma^{+}_H (z)-\sigma^{-}_H (z))/4$, such that $\tilde{\sigma}_H(z)$ falls in the area with a probability of $95\%$. The simulated Hubble parameter data point $H_{sim}(z_i)$ at redshift $z_i$ with the associated uncertainty of $\tilde{\sigma}_H(z_i)$ is computed by
 $H_{\rm sim}(z_i)=H_{\rm fid}(z_i)+\Delta H_i$, where $\Delta H_i$ is determined via $\mathcal{N}(0, \tilde{\sigma}_H(z_i))$.}
%$
%	p(x; \alpha, \lambda) = \frac{\lambda^{\alpha}}{\Gamma(\alpha)}x^{\alpha-1}e^{-\lambda x}
%$
%where $\alpha$ and $\lambda$ are parameters, and the gamma function is:
%$
%	\Gamma(\alpha)=\int_{0}^{\infty}e^{-t}t^{\alpha-1}dt.
%$
%%The assumend distribution of OHD is shown in the middle panel of Fig, \ref{redshift_distribution}. Then,
%%
%%\begin{figure}
%%	\centering
%%	\includegraphics[width=230pt]{./img/redshift_distribution.png}
%%	\caption{Top: the redshift distribution of 32 actual OHD. Middle: the assumed redshift distribution. Bottom: the redshift distribution of 32 simulated OHD.}
%%	\label{redshift_distribution}
%%\end{figure}
%%
%%\begin{figure}
%%	\centering
%%	\includegraphics[width=230pt]{./img/sigma_Hz.png}
%%	\caption{The error of observational $H(z)$ (red dots) along with their linear regression best-fit (red dashed line).}
%%	\label{sigma_Hz}
%%\end{figure}
%
%In the Fig. \ref{sigma_Hz}, we plot errors with respect to the redshift $z$. As expected, the uncertainties tend to increase with the redshift.
%For the generation of the $H(z)$ mock data,
to mimics the observed points.
For the training procedure, we consider the mock $H(z)$ data set with the number of
hidden layers varying from one to three, and eight network models are trained with $2^n$ number
of neurons, where $7 \leq n \leq 14$. %We therefore train a total of twenty-four network models to determine the optimal network configuration.
The network model for reconstructing $H(z)$ is optimized by using mock $H(z)$ data which has the same number as that of the actual OHD; while the training batch size\footnote{%It should be noted that there are three basic concepts in the training of neural networks: Epoch, batch, and iteration. Epoch: A complete training of the model is performed using all the data from the training set.
	Batch: A small part of the samples in the training set are used to update the parameters of the model weights in a back propagation.
	%\footnote{Batch Normalization was proposed by \cite{Ioffe2015}. Since the fact that the distribution of each layer's inputs changes during training as the parameters of the previous layers change, batch normalization layer is implemented before every nonlinear layer, thus allowing us to use much higher learning rates and be less careful about initialization.}
	%Iteration: The process of making a single parameter update to the model using batch data.
} is set to half of the number of the $H(z)$ data points.
%\footnote{%\subsection{The $H(z)$ data}
%The model-independent OHD have unique advantages to calibrate GRBs in a model-independent way \citep{Amati2019}.
%OHD can be obtained with the cosmic chronometers (CC) method, which relates the evolution of differential ages of passive galaxies at different redshifts without assuming any cosmological model \citep{Jimenez2002}.
%\begin{equation}
%	H(z)=-\frac{1}{1+z}\frac{dz}{dt}\,.
%\end{equation}
%Applying the CC approach to the passively evolving galaxies from the luminous red galaxy (LRG) sample, 11 $H(z)$ data in the range $z\leq1.8$ were obtained \citep{Jimenez2003,Simon2005,Stern2010}.
%Based on another analysis, \cite{Moresco2012,Moresco2015,Moresco2016} obtained 15 additional OHD in the range $0.179\leq z\leq1.965$.
%With a full-spectrum fitting technique, \cite{Zhang2014} determined 4 additional estimates of OHD at $z<0.3$, and \cite{Ratsimbazafy2017} obtained one point at $z=0.47$, respectively.
%Recently, the 31 OHD \cite{Jimenez2003,Simon2005,Stern2010,Moresco2012,Moresco2015,Moresco2016,Zhang2014,Ratsimbazafy2017} have been widely used for cosmological purposes \citep{Capozziello2018,Amati2019,Li2020,Montiel2021,Luongo2023,Vagnozzi2021,Dhawan2021,Liu2022b,Liang2022}.
%\cite{Tomasetti2023} proposed a similar approach to obtain a new point  $H(z)=135\pm65$ at $z=1.26$.
We calculate the $\rm RISK$\footnote{Following \cite{WangGJ2020}, we adopt the risk statistic to select the optimal number of hidden layers of the network:
	$\rm RISK =  \sum_{i=1}^{N} \rm BIAS^2_i + \sum_{i=1}^{N} \rm VARIANCE_i,$
	%\begin{equation}
	%	\begin{aligned}
		%		\rm RISK &= 	\displaystyle \sum_{i=1}^{N} \rm BIAS^2_i +
		%		\displaystyle \sum_{i=1}^{N} \rm VARIANCE_i \\
		%		&=
		%		\displaystyle \sum_{i=1}^{N} (H(z_i)-\bar{H}(z_i))^2 + \displaystyle \sum_{i=1}^{N} \sigma^2(H(z_i))
		%	\end{aligned}	
	%\end{equation}
	where $N$ is the number of $H(z)$ data points, $\bar{H}(z_i)$ denotes the fiducial value of $H(z)$.} of eight models for each network structure.

The initial learning rate is set to 0.01 which will decrease with the number of iterations. The network is trained after $3\times10^5$ iterations, ensuring that the function no longer decreases. We inferred that the network model with one hidden layer minimizes the $\rm RISK$ with respect to the two and three hidden layer training networks.
Given that we are using $H(z)$ data alone, the lack of complexity in the data seems to infer a simpler one-layer structure to the ANN.
%We now determine the optimal number of neurons with one hidden layer via the consideration of eight network models with a varying number of neurons.
We %illustrate the $RISK$ values for each network model in Fig. \ref{fig_RISK}, in which one
could clearly observe find that 4096 neurons minimize the risk function.
Consequently, the network structure with one hidden layer and 4096 neurons was found to be the optimal network structure and will therefore be adopted in our $H(z)$ reconstructions.

\begin{figure*}[htbp]
	\centering
	\includegraphics[width=360pt]{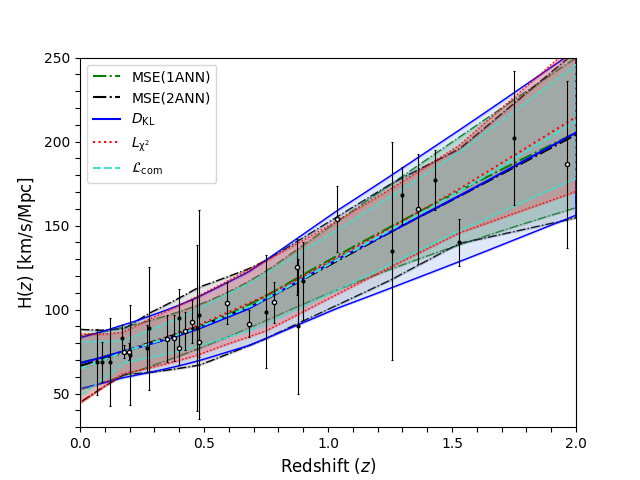}
	\caption{Results of reconstruction used the latest OHD  with different loss functions ($D_{\rm KL}, L_{\chi^2}$, and $\mathcal{L_{\rm com}}$) as well as MSE (one and two ANNs to train $\sigma_{H(z)}$ and $H(z)$ without accounting for the data uncertainties) respectively. The black cycles and dots indicate the 15 OHD with the full covariance matrices and the other 18 OHD without covariance, respectively. The error represent only the diagonal part of the covariance matrix; the full covariance should be taken into account to use the data appropriately.
		%The dashed green curve is obtained with $H_0$ = 67.36 km $s^{-1}$ ${{\rm Mpc}}^{-1}$, $\Omega_m$ = 0.315 \citep{Plank2020}, and the solid green curve is obtained with  $H_0$ = 73.6 km $s^{-1}$ ${{\rm Mpc}}^{-1}$, $\Omega_m$ = 0.334 \citep{Brout2022}.
		%The loss function for each graph is indicated below the corresponding graph.
	}
	\label{fig_Re}
\end{figure*}

In Fig. \ref{fig_Re}, we show results of reconstruction for the latest OHD  with different loss functions ($D_{\rm KL}, L_{\chi^2}$,  $\mathcal{L_{\rm com}}$), as well as MSE (one and two ANNs to train $\sigma_{H(z)}$ and $H(z)$ without accounting for the data uncertainties) respectively. Although the shapes of reconstruction of $H(z)$ are similar,
results with the loss function %that combines the covariance matrix and KL divergence,
$\mathcal{L_{\rm com}}$ exhibits a more accurate and stable reconstruction.
This is particularly evident in the regions where the data points are sparse, $\mathcal{L_{\rm com}}$ appears  capture the underlying trend and minimize fluctuations more accurately.
As a result, $\mathcal{L_{\rm com}}$ demonstrates superior performance in modeling the Hubble parameter $H(z)$, particularly when compared to the $D_{\rm KL}$ and $L_{\chi^2}$ functions, which do not fully leverage the data's intrinsic relationships. %where the black dots and their error bars represent the actual observational Hubble data, and the shaded regions indicate the uncertainties predicted by the neural network.
%Notably, in the reconstruction shown at (c), we empsloyed a network structure with one hidden layer and 4096 neurons with the loss function was MSE, and the output nodes were reduced from two to one. The (d) of Fig.\ref{fig_Re} shows the reconstruction result.
%Here, we also use the network structure with 1 hidden layer and 4096 neurons, and the loss function uses MSE, but the output node is changed from the original two to one. %Fig. \ref{fig_Re_Zhang2024} shows the $H(z)$ reconstructed by ANN of our OHD data.

\section{GRB Hubble Diagram}
%The Amati relation \citep{Amati2002}  connects the spectral peak energy ($E_{\rm p}$) and the isotropic equivalent radiated energy ($E_{\rm iso}$), which can be expressed as
%\begin{equation}
%	y = a + bx
%\end{equation}
%where $y \equiv \log_{10}\frac{E_{\rm iso}}{1{\rm erg}},\quad x \equiv \log_{10}\frac{E_p}{300{\rm keV}}$,
%$a$ and $b$ are free coefficients needing to be calibrated from the observed GRBs data in the formula. $E_{\rm iso}$ and $E_{\rm p}$ can be respectively expressed as:
%\begin{equation}
%	E_{{\rm iso}} = 4\pi d^2_L(z)S_{{\rm bolo}}(1+z)^{-1},\quad E_p = E^{{\rm obs}}_p(1+z)
%\end{equation}
%In this paper, we calibrate the Amati relation  with J220 GRB data \citep{Jia2022} \footnote{Following \cite{Khadka2021,Cao2024}, we have removed GRB020127 which has an unreliable redshift from the 221 GRB sample \citep{Jia2022}.} at low redshift by reconstructing the Hubble parameter from the latest OHD in ANN framework with  $\mathcal{L_{\rm com}}$ as the loss function.
In this paper, we calibrate the Amati relation  with A219 GRB data\footnote{Following \cite{Liang2022}, we excluded GRB051109A from the A220 sample, which incorporates the A118 dataset with the lowest intrinsic dispersion, along with the A102 dataset from the 193 GRBs analyzed by \cite{Amati2019} and \cite{Demianski2017}.} and J220 GRB data\footnote{Following \cite{Khadka2021} and \cite{Cao2024}, we have removed GRB020127 which has an unreliable redshift from the J221 GRB sample \citep{Jia2022}.} at low redshift by reconstructing the Hubble parameter from the latest OHD in ANN framework to build GRB Hubble diagram. % with  $\mathcal{L_{\rm com}}$ as the loss function.
For reconstruction, we use the recent 33 point at $0.179\leq z\leq1.965$ including the new points at $z=0.80$ \citep{Jiao2022} and at $z=1.26$ \citep{Tomasetti2023}. Therefore, we use the redshift cutoff at $z=1.965$ to calibrate the Amati relation with  A219 (115 GRBs) and J220 (128 GRBs) samples. \cite{Montiel2021} calibrated the Amati relation with OHD at $z<1.43$, according to the analysis of \cite{Moresco2020} that OHD was carried out through simulations in the redshift range
$0 < z < 1.5$. %we limit our Hubble data to this range and instead of using the 31 measurements of $H(z)$ used in \citep{Capozziello2018}.
In order to compare with the previous analyses \citep{Liang2022,Liu2022b,LZL2023}, we also use a subsample including 29 OHD points with a redshift cutoff at $z=1.4$ to calibrate the Amati relation with  A219 (79 GRBs) and J220 (90 GRBs) samples.

The Amati relation \citep{Amati2002}, which refers to the power-law relationship between the spectral peak energy $E_{\text{peak}}$ and the isotropic equivalent energy $E_{\text{iso}}$ of GRBs, %indicating that the two are directly proportional. % Mathematically, this relationship
can be expressed %as $ E_{\text{peak}} \propto E_{\text{iso}} $. or
in the logarithmic form,
\begin{equation}
	\log E_{\text{iso}} = a + b \cdot \log E_{\text{p}}
\end{equation}
where $ a $ and $ b $ are free coefficients, %that need to be calibrated based on observed GRB data.
and
\begin{equation}
	E_{{\rm iso}} = 4\pi d^2_L(z)S_{{\rm bolo}}(1+z)^{-1},\quad E_\text{p} = E^{{\rm obs}}_\text{p}(1+z)
\end{equation}
%where $E^{\rm obs}_p$ is the observational value of GRB spectral peak energy and $S_{\rm bolo}$ is observational value of bolometric fluence, both $E^{{\rm obs}}_p$ and $S_{{\rm bolo}}$ can be observable.
%The luminosity distance can be calculated by the reconstructed $H(z)$ at the redshift of GRBs
%\footnote{
	%	For the recent Planck results \citep{Plank2020} favor a flat spatial curvature, we consider a flat space in this work. However, recently works constrain non-spatially flat models with GRBs and results are promising \citep{Luongo2023}.
	%}
%\begin{eqnarray}
%	d^{\rm ANN}_{L} = c(1+z)\int^z_0\frac{dz^{'}}{H(z^{'})}
%\end{eqnarray}
where \( E_{\text{iso}} \) is %the isotropic equivalent radiated energy measured
in units of 1 erg, \( E^{\text{obs}}_\text{p} \) %represents the observed value of the spectral peak energy measured
in units of 300keV, \( S_{\text{bolo}} \) is the observed value of the bolometric fluence. %Both \( E^{\text{obs}}_p \) and \( S_{\text{bolo}} \) are observable.
%here the value of $H(z)$ at redshift $z$ can be reconstructed with OHD by ANN at the redshift of GRBs.
The values of the Hubble parameter $H(z)$ can be reconstructed by ANN specifically at redshifts of GRBs.
For a flat space, the luminosity distance can be calculated, %using the reconstructed \( H(z) \),
\begin{eqnarray}
	d^{\rm ANN}_{L} = c(1+z)\int^z_0\frac{dz^{'}}{H(z^{'})}.
\end{eqnarray}

It is important to note that the likelihood approach by \cite{D'Agostini2005}
could introduce subjectivity regarding the selection of the independent variable. %To mitigate this, \cite{Liang2008} adopted the method of the bisector of the two ordinary least-squares, as also used in \cite{Schaefer2007}.
The likelihood function proposed by \cite{Reichart2001} which advantageously eliminates the need to arbitrarily designate an independent variable between $ E_{p} $ and $ E_{\text{iso}} $  has been used in \citep{Amati2013}.
The \cite{D'Agostini2005} and \cite{Reichart2001} likelihood functions can be expressed as \citep{Lin2016,LZL2023}
\begin{eqnarray}
	\mathcal{L}_{\rm D}\propto\prod_{i=1}^{N_1} \frac{1}{\sigma_{\rm D}}
	\times\exp\left[-\frac{[y_i-y(x_i,z_i; a, 	b)]^2}{2\sigma^2_{\rm D}}\right],
\end{eqnarray}
here $\sigma_{\rm D}=\sqrt{\sigma_{\rm int, D}^2 + \sigma_{y_i}^2 + b^2\sigma_{x_i}^2}$, the intrinsic scatter $\sigma_{\rm int, D}$ which represents any other unknown errors except for the measurement error can be treated as a free parameter. %in fitting procedure.
\begin{eqnarray}
	\mathcal{L}_{\rm R}\propto\prod_{i=1}^{N_1} \frac{\sqrt{1+b^2}}{\sigma_{\rm R}}
	\times\exp\left[-\frac{[y_i-y(x_i,z_i; a, b)]^2}{2\sigma^2_{\rm R}}\right].
\end{eqnarray}
here $\sigma_{\rm R}=\sqrt{\sigma_{\rm int, R}^2 + \sigma_{y_i}^2 + b^2\sigma_{x_i}^2}$,  $\sigma_{\rm int, R}$ can be calculated by the "equivalent" total intrinsic scatter  \citep{Lin2016}: $\sigma_{\rm equi}=\sqrt{\sigma_{y,\rm ext}^2 + b^2\sigma_{x,\rm ext}^2}$, in which $\sigma_{x,\rm ext}$ and $\sigma_{y,\rm ext}$ are the extrinsic scatter along the $x$-axis and $y$-axis \citep{Reichart2001}. In this work, we treat the total intrinsic scatter $\sigma_{\rm int, R}$ as a free parameter.

We employ the Markov Chain Monte Carlo (MCMC) method, facilitated by the Python package \textit{emcee} \citep{ForemanMackey2013}, which refines the Metropolis-Hastings algorithm.
The calibration outcomes, encompassing the intercept $ a $, the slope $ b $, and the intrinsic scatter $ \sigma_{\text{int}} $, for the A219 and J220 samples at $z\leq1.965$ and $z\leq1.4$ are detailed in Table \ref{Amati relation}.
%\textbf{as well as $\sigma_{\rm int}$ fitted by $\sigma_{\rm int}=\sqrt{\sigma_{y,\rm int}^2 + b^2\sigma_{x,\rm int}^2}$  are smaller than ones treated as a free parameter by the \cite{D'Agostini2005} method and the \cite{Reichart2001} method with GRBs.}
We find that the values of the intercept ($a$) by the two likelihood function methods are consistent in 1$\sigma$ uncertainty; and
the values of slope ($b$) by \cite{Reichart2001} method are close to the typical value ($b=2, E_{\rm iso} \propto E_{\rm p}^2$);
the value of the intrinsic scatter of the \cite{Reichart2001} method calculated by the fitting values of $\sigma_{y,\rm int}$ and $\sigma_{x,\rm int}$  are smaller than the total intrinsic scatter $\sigma_{\rm int, R}$ treated as a free parameter, which indicate that there are possible systematic errors %which are not including $\sigma_{y,\rm int}$ and $\sigma_{x,\rm int}$  of the intrinsic scatter
in fitting procedure.
We also find the calibration results with the GRB samples at $z\leq1.965$ are consistent with the GRB samples at $z\leq1.4$. For the A219 sample at $ z < 1.4 $, we find the fitting value of slope ($b=1.28^{+0.13}_{-0.13}$) by \cite{D'Agostini2005} method
align with prior analyses which based on SNe Ia at $ z < 1.4 $ utilized GaPP: $1.298^{+0.090}_{-0.080}$ \citep{Liang2022}  and interpolation: $1.290^{+0.126}_{-0.126}$ \citep{Liu2022b}. %The results for the J220 align with prior analyses the fitting results of the fourth sub-sample with the small extrinsic scatter
For the J220 sample\footnote{\cite{Jia2022} utilized the standard cosmological parameters from \cite{Plank2020} ($\Omega_{\rm m}$ = 0.315, $\Omega_{\Lambda}$ = 0.685, and $H_0$ = 67.4 km $s^{-1}$ ${{\rm Mpc}}^{-1}$) without any error to obtain the values of $E_{{\rm iso}}$. \cite{Cao2024} used the updated values for 7 GRBs from \cite{Dirirsa2019} by simultaneously constraining. %correlation parameters and cosmological parameters.
Here we use the original values fitted with the Band model \citep{Jia2022} in J220 GRB sample. %for reconstruction from ANN. %GaPP and the simultaneous fitting method.
} at $ z < 1.4 $ our result  ($b=1.57^{+0.09}_{-0.09}$) is consistent with that of the fourth sub-sample ($1.35\le z<1.49$) in \cite{Jia2022}: %in which the extrinsic scatter sub-sample is small
 $1.51^{+0.17}_{-0.17}$ and that by GaPP: $1.59^{+0.10}_{-0.10}$ \citep{Xie2023}, %as well as Plank Collaboration are
which is different with \cite{Cao2024} by the simultaneous fitting for the $\Lambda$CDM model: $1.171^{+0.087}_{-0.087}$.
It is worth noting that for the values of $\sigma_{\rm int}$ obtained with the J220 sample  are smaller than ones obtained with the A219 sample; and the values of $\sigma_{\rm int}$ obtained by the \cite{D'Agostini2005} method are smaller than ones obtained by the \cite{Reichart2001} method.

\begin{table*}
	\centering
	\caption{
		Calibration results (the intercept $a$, the slope $b$, and the intrinsic scatter $\sigma_{\rm int}$) with the Amati relation for the A219 and J220 samples at $z\leq1.965$ and $z\leq1.4$.% by the likelihood method \citep{Reichart2001}  and the likelihood method \citep{D'Agostini2005}.
	}
	\label{Amati relation}
	\begin{tabular}{lcccc}
		%		\toprule
		%		Methods & $a$ & $b$ & $\sigma_{{\rm int}}$ \\
		%		\hline
		%		\specialrule{0em}{1pt}{1pt}
		%		\cite{D'Agostini2005} & $52.727^{+0.061}_{-0.061}$ & $1.28^{+0.13}_{-0.13}$ &
		%		$0.527^{+0.038}_{-0.050}$ \\
		%		\cite{Reichart2001} & $52.748^{+0.075}_{-0.075}$ &
		%		$2.02^{+0.16}_{-0.23}$ &
		%		$0.50^{+0.29}_{-0.29}$ \\
		%		\bottomrule
		\toprule
		Sample&Methods & $a$ & $b$ & $\sigma_{{\rm int, D}}/\sigma_{\rm int, R}$
		\\
		\hline
		\specialrule{0em}{1pt}{1pt}
		A219 (115 GRBs at $z\leq1.965$)&D'Agostini & $52.79^{+0.05}_{-0.05}$ & $1.34^{+0.10}_{-0.10}$ &
		$0.52^{+0.03}_{-0.04}$ \\
		&Reichart & $52.77^{+0.06}_{-0.06}$ &
		$1.90^{+0.01}_{-0.03}$ &
		$0.58^{+0.04}_{-0.05}$
		\\
		\hline
		\specialrule{0em}{1pt}{1pt}
		J220 (128 GRBs at $z\leq1.965$)&D'Agostini & $52.84^{+0.04}_{-0.04}$ & $1.52^{+0.08}_{-0.08}$ &
		$0.42^{+0.03}_{-0.03}$
		\\
		&Reichart & $52.83^{+0.04}_{-0.04}$ &
		$1.90^{+0.09}_{-0.03}$ &
		$0.45^{+0.03}_{-0.04}$
		\\
		\hline

		\specialrule{0em}{1pt}{1pt}
		A219 (79 GRBs at $z\leq1.4$)&D'Agostini & $52.73^{+0.06}_{-0.06}$ & $1.28^{+0.13}_{-0.13}$ &
		$0.53^{+0.04}_{-0.05}$ \\
		 &Reichart & $52.75^{+0.07}_{-0.07}$ &
		$1.86^{+0.13}_{-0.05}$ &
		$0.59^{+0.05}_{-0.06}$
		\\
		\hline
		\specialrule{0em}{1pt}{1pt}
		J220 (90 GRBs at $z\leq1.4$)&D'Agostini & $52.80^{+0.05}_{-0.05}$ & $1.57^{+0.09}_{-0.09}$ &
		$0.41^{+0.03}_{-0.04}$
		\\
		 &Reichart & $52.81^{+0.05}_{-0.05}$ &
		$1.89^{+0.11}_{-0.03}$ &
		$0.43^{+0.15}_{-0.08}$
		\\
		\bottomrule
	\end{tabular}
\end{table*}

\begin{figure*}
	\centering
	\begin{subfigure}[b]{0.47\textwidth}
		\centering
		\includegraphics[width=\textwidth]{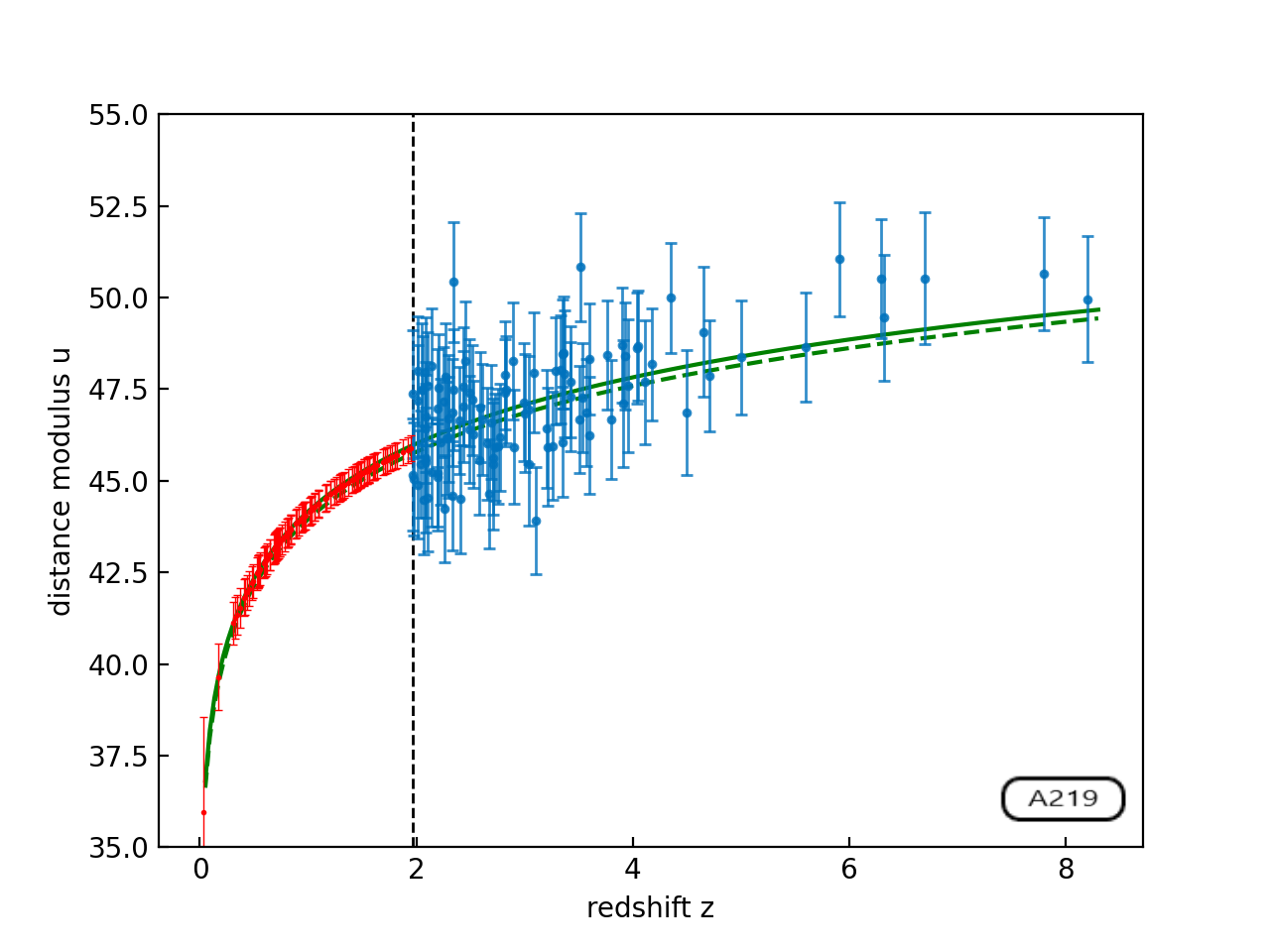}
		
	\end{subfigure}
	\begin{subfigure}[b]{0.47\textwidth}
		\centering
		\includegraphics[width=\textwidth]{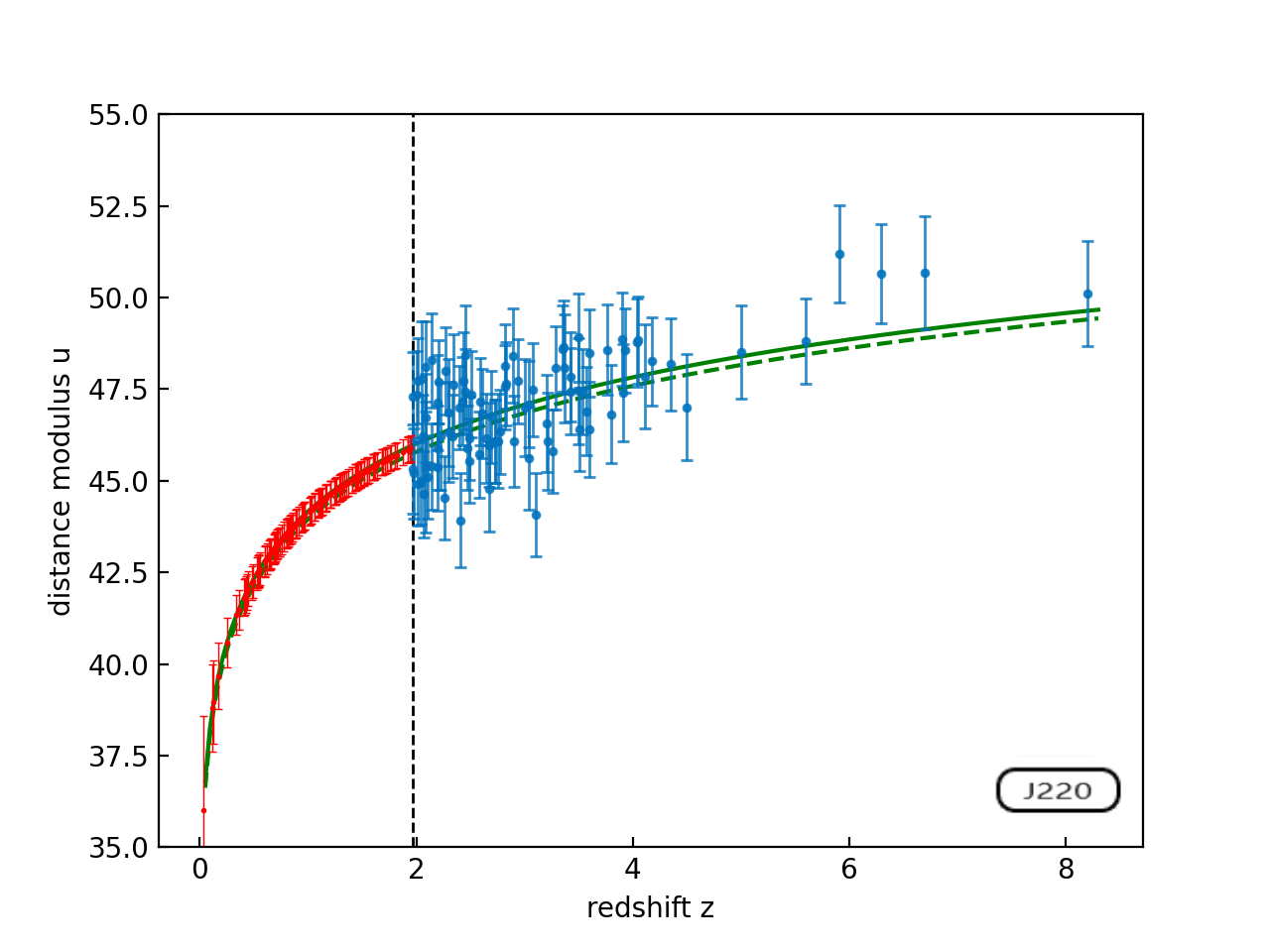}
		
	\end{subfigure}
	\caption{
		%		GRB Hubble diagram with the J221 data set. GRBs at $z < 1.4$ were obtained by ANN from the OHD data (purple points), while GRBs with $z > 1.4$ (blue points) were obtained by the Amati relation and calibrated with A118 at $z < 1.4$. The solid green curve is the CMB standard distance modulus with $H_0$ = 67.36 km $s^{-1}$ ${{\rm Mpc}}^{-1}$, $\Omega_m$ = 0.315 \citep{Plank2020}, and the green long dotted curve is the SNIa standard distance modulus with $H_0$ = 74.3 km $s^{-1}$ ${{\rm Mpc}}^{-1}$, $\Omega_m$ = 0.298 \citep{Scolnic2018}. The black dotted line denotes $z = 1.4$.
		%		
		The Hubble diagram of GRBs is presented with the A219 (\emph{left}) and J220 (\emph{right}) samples. GRBs data at $z \le 1.965$ (\emph{red points}) were obtained from the OHD data using ANN, GRBs data at $z>1.965$ (\emph{blue points}) are derived from the Amati relation. %and calibrated with the GRB data \textbf{at $z < 1.975$}.
The solid curve and the dotted curve  represent the standard distance modulus fitting from CMB: $H_0$ = 67.36 km $s^{-1}$ ${{\rm Mpc}}^{-1}$, $\Omega_m$ = 0.315 \citep{Plank2020}, and %the standard distance modulus fitting from
Pantheon+:  $H_0$ = 73.6 km $s^{-1}$ ${{\rm Mpc}}^{-1}$, $\Omega_m$ = 0.334 \citep{Brout2022}, respectively.
	}
	\label{fig_HD}
\end{figure*}

%Our analysis indicates that the likelihood method proposed by Reichart yields a larger slope $b$ compared to the method by D'Agostini.
To ensure objectivity in selecting the independent variable, we choose to use the calibration results by Reichart's likelihood method to construct the GRB Hubble diagram. % for redshifts greater than 1.4. %This approach facilitates a more impartial assessment and interpretation of the observational data. %Assuming the calibration results of the Amati relation at $z<1.4$ are valid at high redshift, we can derive the luminosity distances of GRBs at $z>1.4$. The GRB Hubble diagram is plotted in Figure \ref{fig_HD}. The uncertainty of GRB distance modulus with the Amati relation is
Assuming that the calibration results of the Amati relation at $z \le 1.965$ are still applicable at higher redshifts, we can deduce the luminosity distances of GRBs at $z > 1.965$. The Hubble diagram for GRBs is illustrated in Figure \ref{fig_HD}. The uncertainty in the GRB distance modulus is
\begin{eqnarray}
	\sigma^{2}_{\mu} = (\frac{5}{2}\sigma_{\log_{\frac{E_{{\rm iso}}}{1{\rm erg}}}})^{2} + (\frac{5}{2\ln10}\frac{\sigma_{S_{{\rm bolo}}}}{S_{{\rm bolo}}})^{2}
\end{eqnarray}
where $
%\begin{eqnarray}\label{eqnarray7}
\sigma^{2}_{\log_{\frac{E_{\rm iso}}{1{\rm erg}}}} = \sigma^{2}_{\rm int} + (\frac{b}{\ln10}\frac{\sigma_{E_{p}}}{E_{p}})^{2} + \sum \bigg (\frac{\partial_{y}(x;\theta_c)}{\partial \theta_{c_{i}}} \bigg)^2C_{ii}
%\end{eqnarray}
$,
here $\theta_c = {\sigma_{{\rm int}}, a, b}$, and $C_{ii}$ means the diagonal element of the covariance matrix of these fitting coefficients.

\section{Constraints on DE Models}
In order to constrain cosmological models, we use the GRB data in the Hubble diagram at $z>1.4$ with the Pantheon+ sample \citep{Scolnic2022}, which consists of 1701 light curves of 1550 unique spectroscopically confirmed SNe Ia at the redshift range $0.001<z<2.26$ with the observed distance modulus of SNe given by
\begin{equation}
	\mu_{\rm SN} = m_{\rm B,corr}^* - M_{\rm B},
	\label{eq:muSN}
\end{equation}
where $m_{\rm B,corr}^* = m_{\rm B}^* + \alpha X_1 -\beta \mathcal{C}+ \Delta_M + \Delta_B$,	
$m_{\rm B}^*$ is the observed peak magnitude in rest frame B-band, $X_1$ is the time stretching of the light-curve, $\mathcal{C}$ is the SNe color at maximum brightness; $\alpha, \beta$  are nuisance parameters which should be fitted simultaneously with the cosmological parameters,
$\Delta_M$, $\Delta_B$ are the distance correction based on the host galaxy mass of the SN and  distance
correction based on predicted biases from simulations, respectively; $M_{\rm B}$ is the absolute magnitude which can be calibrated by setting an absolute distance scale with primary distance anchors such as Cepheids.
The SH0ES (Supernovae and $H_0$ for the Equation of State of dark energy) team obtained $M_B = -19.253\pm0.027$ \citep{Riess2022} from the Pantheon+ sample \citep{Scolnic2022, Brout2022}.
In this work, we set $M_{\rm B} = -19.253$ \citep{Riess2022} for simplicity. % to obtain the distance modulus of the Pantheon+ sample.
\cite{Liu2022b}  used $M_{\rm B}=-19.36$ with the Pantheon sample, BAO, and CC \citep{Gomez-Valent2022} to calibrate the Amati relation and the improved Amati relation.
\cite{Mukherjee2024a} employed neural networks to obtain: $M_{\rm B} =-19.353^{+0.073}_{-0.078}$ from the Pantheon+ and CC sample. %Following \cite{Mukherjee2024a},
\cite{Mukherjee2024b} calibrated the Dainotti relations by setting $M_{\rm B} =-19.35$  for simplicity. In order to investigate the influence for the value of $M_{\rm B}$ in the subsequent constraints,
we also use $M_{\rm B} = -19.353$ ($\Delta M_B\approx0.10$) for comparison. It should be noted that $M_{\rm B}$ can be also treated as a free parameter in fitting procedure \citep{Liang2022,LZL2023};
and we can also treat $M_{\rm B}-\mu_0$ (where $\mu_0=5\log h-42.38$, $h=H_0/(100{\rm km/s/Mpc})$, $H_0$ is the Hubble constant) as a free parameter, which can be marginalized over analytically to avoid the degeneracy between $M_B$ and $H_0$ \citep{Li2020}.

The $\chi^2$ for the distance modulus can be expressed as
\begin{equation}\chi^2_{\mu} = \sum^{N}_{i=1} \left[\frac{\mu_{\rm obs}(z_i)-\mu_{\rm th}(z_i;p,H_0)}{\sigma_{\mu_i}}\right]^2.
\end{equation}
Here $\mu_{{\rm obs}}$ is the observational value of distance modulus and its error $\sigma_{\mu_i}$, and $\mu_{{\rm th}}$ is the theoretical value of distance modulus calculated from the cosmological parameters $p$,
%which can be calculated as
\begin{eqnarray}\label{mu}
	\mu_{\rm th}=5\log \frac{d_L}{\textrm{Mpc}} + 25=5\log D_L-\mu_0,
\end{eqnarray}
%where $\mu_0=5\log h-42.385$, $h=H_0/(100{\rm km/s/Mpc})$, $H_0$ is the Hubble constant.
For a flat space, the unanchored luminosity
distance $D_L$ can be calculated by
$
D_L\equiv
\frac{H_0d_L}{c}
=(1+z)\int_0^z\frac{dz'}{E(z')},
$
where $c$ is the light speed,  $E(z)=[\Omega_{\textrm{M}}(1+z)^3+\Omega_{\textrm{DE}}X(z)]^{1/2}$, and  $X(z)=\exp[3\int_0^z\frac{1+w(z')}{1+z'}dz']$, which is determined by the choice of DE model.

%Constraints only with 131 GRBs (J221) $z > 1.4$ are shown in Fig. \ref{fig_GRB_SN_LCDM_N} ($\Lambda$CDM), Fig. \ref{fig_GRB_SN_wCDM_N} ($w$CDM) and Fig. \ref{fig_GRB_SN_CPL_N} (CPL), which are summarized in Table \ref{joint constraint results}.	
%
%	
%	\begin{figure}
	%		\centering
	%		\includegraphics[width=0.35\linewidth]{./img/GRB_LCDM_N.png}
	%		\caption{
		%			Constraints on parameters of $\Omega_m$, $h$ for the flat $\Lambda$CDM model with 131 GRBs at $z > 1.4$.
		%		}
	%		\label{fig_GRB_LCDM_N}
	%	\end{figure}
%	
%	\begin{figure}
	%		\centering
	%		\includegraphics[width=0.35\linewidth]{./img/GRB_wCDM_N.png}
	%		\caption{
		%			Constraints on parameters of $\Omega_m$, $h$, and $w_0$  for the flat $w$CDM model with 131 GRBs at $z > 1.4$.
		%		}
	%		\label{fig_GRB_wCDM_N}
	%	\end{figure}
%	
%	\begin{figure}
	%		\centering
	%		\includegraphics[width=0.35\linewidth]{./img/GRB_CPL_N.png}
	%		\caption{
		%			Constraints on parameters of $\Omega_m$, $h$, $w_0$ and $w_a$  for the flat CPL model  with 131 GRBs at $z > 1.4$.
		%		}
	%		\label{fig_GRB_CPL_N}
	%	\end{figure}

BAOs offer a distinct perspective on the universe's structure and evolution, which can be used as an invaluable tool for probing cosmological models  at mid-redshift.
In order to %refine our analysis and
tighten the constraints on cosmological parameters, we combine  GRBs and SNe with BAOs.\footnote{It should be noted that BAO measurements which under a fiducial cosmology could provide biased constraints.}
The likelihood of BAO can be expressed as,
\[
\chi_{\mathrm{BAO}}^2 = \Delta P_{\mathrm{BAO}} C_{\mathrm{BAO}}^{-1} \Delta P_{\mathrm{BAO}}^{\mathrm{T}},
\]
where  %\footnote{For uncorrelated points the covariance matrix is a diagonal matrix, and its elements are the inverse errors, and for correlated points, the covariance matrices are from \cite{2021PhRvD.103h3533A}.},
$\Delta P_{\mathrm{BAO}}=v_{\mathrm{obs}}(z) - v_{\mathrm{th}}(z)$, $v_{\mathrm{obs}}(z)$ is a BAO measurement \footnote{The BAO feature appears in both the line-of-sight direction and the transverse direction and provides  measurements of the radial projection: $\frac{D_{\mathrm{H}}(z)}{r_{\mathrm{d}}}=\frac{c}{H(z)r_{\mathrm{d}}}$ and the transverse comoving distance $\frac{D_{\mathrm{M}}(z)}{r_{\mathrm{d}}}=\frac{c}{H_0 r_{\mathrm{d}}}\Gamma(z)$, where $\Gamma(z)=\int_{0}^{z}dz^{'}/E(z^{'})$ in a flat space, the sound horizon is defined as $r_{\mathrm{d}}=\frac{c}{H_0} \int_{z_d}^{\infty} \frac{c_{\mathrm{s}}}{E(z^{'}) dz^{'}}$ at the drag epoch $z_{\mathrm{d}}$.
	%$z_{\mathrm{d}}= \frac{1345 \omega_m^{0.251}}{1+ 0.659 \omega_m^{0.828}}[1+b_1 \omega_b^{b_2}]$, $b_1=0.313 \omega_m^{-0.419}[1+0.607 \omega_m^{0.674}],b_2=0.238 \omega_m^{0.223}$ with $\omega_b=\Omega_{\mathrm{b}} h^2$ and $\omega_m=\Omega_{\mathrm{m}} h^2$;
	%\footnote{The comoving sound horizon \( r_{\mathrm{s}}(z) \) is given as \cite{1998ApJ...496..605E}:
		%$r_{\mathrm{s}} (z)=\frac{c}{H_0} \int_{z}^{\infty} \frac{c_{\mathrm{s}}}{E(z^{'}) dz^{'}}$. The redshift of the drag epoch can be approximated as \cite{1996ApJ...471..542H}: $z_ {\mathrm{d}} = \frac{1345 \omega_m^{0.251}}{1+ 0.659 \omega_m^{0.828}}[1+b_1 \omega_b^{b_2}]$, where $b_1=0.313 \omega_m^{-0.419}[1+0.607 \omega_m^{0.674}],b_2=0.238 \omega_m^{0.223}$ with $\omega_b=\Omega_{\mathrm{b}} h^2$ and $\omega_m=\Omega_{\mathrm{m}} h^2$.},
 We use the constrained value from \cite{Plank2020} ($r_{\mathrm{d}}=147.05\pm 0.30 \rm{Mpc}$, $z_d=1089.80\pm0.21$).   $r_{\mathrm{d}}$ can be also treated as a free parameter \citep{Li2020}. %We set $r_{\mathrm{d}}=147.05 \rm{Mpc}$ in our analysis.

	The angular diameter distance $D_{\mathrm{A}}(z)$ has relation with $D_{\mathrm{M}} (z)$: $D_{\mathrm{A}}(z)=D_{\mathrm{M}}(z)/(1+z)$.} of the observed points at each $z$, and $v_{\mathrm{th}}(z)$ is the prediction of the theoretical model, $C_{\mathrm{BAO}}$ is the covariance matrix of the observed points.
Here we use BAO data including the 6dF Galaxy Survey (6dFGS) at $z_{\rm eff}=0.106$, the Sloan Digital Sky Survey (SDSS) DR7 Main Galaxy Sample (MGS) at $z_{\rm eff}=0.15$, and nine measurements from the extended Baryon Oscillation Spectroscopic Survey (eBOSS) DR16 at $z_{\rm eff}=0.38,0.51,0.70,0.85,1.48$ \citep{Wang2024}. We also consider the latest BAO measurements (${D_{\mathrm{H}}(z)}/{r_{\mathrm{d}}}$, ${D_{\mathrm{M}}(z)}/{r_{\mathrm{d}}}$  and ${D_{\mathrm{V}}(z)}/{r_{\mathrm{d}}}=\frac{1}{r_{\mathrm{d}}}[\frac{cz}{H(z)}]^{1/3}[\frac{D_{\mathrm{L}}}{1+z}]^{2/3}$) from Dark Energy Spectroscopy Instrument (DESI) at $z_{\rm eff}=0.295,0.510,0.706,0.930,1.317,1.491,2.330$ \citep{DESI2024}.

The total $\chi^2$ with the joint data of GRB+SNe and BAO can be expressed as
$\chi^2_{{\rm total}} = \chi^2_{{\rm GRB}} + \chi^2_{{\rm SN}} + \chi_{\mathrm{BAO}}^2$. 	
The python package $emcee$  by the MCMC method \citep{ForemanMackey2013} is used to constrain DE models. We consider three DE models in a flat space, the $\Lambda$CDM model with dark energy EoS $w=-1$ , the $w$CDM model  with a constant Equation of State (EoS),  and the Chevallier-Polarski-Linder (CPL) model \citep{CP2001,Linder2003}  in which DE evolving with redshift as a parametrization  EoS,  $w=w_0+w_az/(1+z)$.

%	\begin{figure}
%		\centering
%		\includegraphics[width=0.35\linewidth]{./img/GRB_LCDM_N.png}
%		\caption{
%			Constraints on parameters of $\Omega_m$, $h$ for the flat $\Lambda$CDM model with 131 GRBs at $z > 1.4$.
%		}
%		\label{fig_GRB_LCDM_N}
%	\end{figure}
%	
%	\begin{figure}
%		\centering
%		\includegraphics[width=0.35\linewidth]{./img/GRB_wCDM_N.png}
%		\caption{
%			Constraints on parameters of $\Omega_m$, $h$, and $w_0$  for the flat $w$CDM model with 131 GRBs at $z > 1.4$.
%		}
%		\label{fig_GRB_wCDM_N}
%	\end{figure}
%	
%	\begin{figure}
%		\centering
%		\includegraphics[width=0.35\linewidth]{./img/GRB_CPL_N.png}
%		\caption{
%			Constraints on parameters of $\Omega_m$, $h$, $w_0$ and $w_a$  for the flat CPL model  with 131 GRBs at $z > 1.4$.
%		}
%		\label{fig_GRB_CPL_N}
%	\end{figure}
\begin{table*}
	\centering
	\setlength{\tabcolsep}{2pt} % 减小列间距
	\renewcommand{\arraystretch}{1.4} % 调整行间距
	\caption{
		Constraints on parameters of $\Omega_m$, $h$, $w_0$ and $w_a$ for the flat $\Lambda$CDM model, the flat $w$CDM model, and the flat CPL model only with the A219  and J220  samples ($z>1.965$).
	}
	\label{GRB constraint results}
	\begin{tabular}{cccccccccc}
		\toprule
		Models & Data Set & $\Omega_{m}$ & $h$ & $w_0$ & $w_a$ & $\Delta$AIC & $\Delta$BIC
		%			$-2\ln \mathcal{L_{\rm R}}$  & $\Delta \rm AIC$  & $\Delta \rm BIC$
		\\
		\hline
		\specialrule{0em}{1pt}{1pt}
		$\Lambda$CDM  &
		A219  &
		$0.377^{+0.071}_{-0.17}$ &
		$0.629^{+0.056}_{-0.082}$  &
		- &
		- &
		0 &
		0
		\\
		$w$CDM  &
		A219  &
		$0.380^{+0.11}_{-0.17}$ &
		$0.620^{+0.041}_{-0.11}$  &
		$-1.09^{+0.58}_{-0.76}$ &
		- &
		1.82 &
		4.46
		\\
		CPL &
		A219   &
		$0.382^{+0.070}_{-0.17}$  &
		$0.630^{+0.051}_{-0.10}$  &
		$-1.02^{+0.55}_{-0.55}$ &
		$-1.01^{+0.58}_{-0.58}$ &
		3.58 &
		8.87
		\\
		\hline
		\specialrule{0em}{1pt}{1pt}
		$\Lambda$CDM  &
		J220  &
		$0.379^{+0.072}_{-0.17}$ &
		$0.633^{+0.059}_{-0.080}$  &
		- &
		- &
		0 &
		0
		\\
		$w$CDM  &
		J220  &
		$0.380^{+0.084}_{-0.17}$ &
		$0.621^{+0.044}_{-0.11}$  &
		$-1.08^{+0.51}_{-0.51}$ &
		- &
		1.78 &
		4.30
		\\
		CPL &
		J220   &
		$0.386^{+0.076}_{-0.17}$  &
		$0.628^{+0.045}_{-0.10}$  &
		$-0.96^{+0.88}_{-0.37}$ &
		$-1.03^{+0.58}_{-0.58}$ &
		3.44 &
		8.49
		\\
		\hline
		\bottomrule
	\end{tabular}
\end{table*}

Constraints only with A219 and J220 samples at $z > 1.965$ are %shown in Fig. \ref{fig_GRB_SN_LCDM_N} ($\Lambda$CDM), Fig. \ref{fig_GRB_SN_wCDM_N} ($w$CDM) and Fig. \ref{fig_GRB_SN_CPL_N} (CPL), which are
summarized in Table \ref{GRB constraint results}.
%Due to the degeneracy with the correlation intercept parameter, GRB data alone cannot constrain $H_0$. We follow previous works \citep{Khadka2021, Liang2022} and fix $H_0$=$70\ {\rm km}\ {\rm s}^{-1}{\rm Mpc}^{-1}$ for the case of GRBs alone.
For the $\Lambda$CDM model, we obtain $\Omega_{\rm m}=0.377^{+0.071}_{-0.17}$, $h=0.629^{+0.065}_{-0.082}$, with the A219 sample, and  $\Omega_{\rm m}=0.379^{+0.072}_{-0.17}$, $h=0.633^{+0.059}_{-0.080}$,  with the J220 sample at the 1$\sigma$ confidence level, respectively.	There is evident that the $\Lambda$CDM model ($w_0=-1$, $w_a=0$) remains the best-fit to the observational data according to the results of the $w$CDM model and the CPL model.
We find the results using GRBs alone with the A219 sample are consistent with the J220 sample, which are compatible with the previous works of \cite{LZL2023}.

\begin{figure*}
		\centering
		%		\includegraphics[width=0.35\linewidth]{./img/GRB_SN_BAO_LCDM.png}
%		\begin{subfigure}[b]{0.4\textwidth}
%			\centering
%			\includegraphics[width=\textwidth]{./img/constrain/LCDM_A219_SN_SDSS.png}
%		\end{subfigure}
%		\begin{subfigure}[b]{0.4\textwidth}
%			\centering
%			\includegraphics[width=\textwidth]{./img/constrain/LCDM_J220_SN_SDSS.png}
%		\end{subfigure}
%		\begin{subfigure}[b]{0.4\textwidth}
%	        \centering
%	        \includegraphics[width=\textwidth]{./img/constrain/LCDM_A219_SN_DESI.png}
%        \end{subfigure}
%        \begin{subfigure}[b]{0.4\textwidth}
%	        \centering
%	        \includegraphics[width=\textwidth]{./img/constrain/LCDM_J220_SN_DESI.png}
%        \end{subfigure}		
		\begin{subfigure}[b]{0.4\textwidth}
		        \centering
		        \includegraphics[width=\textwidth]{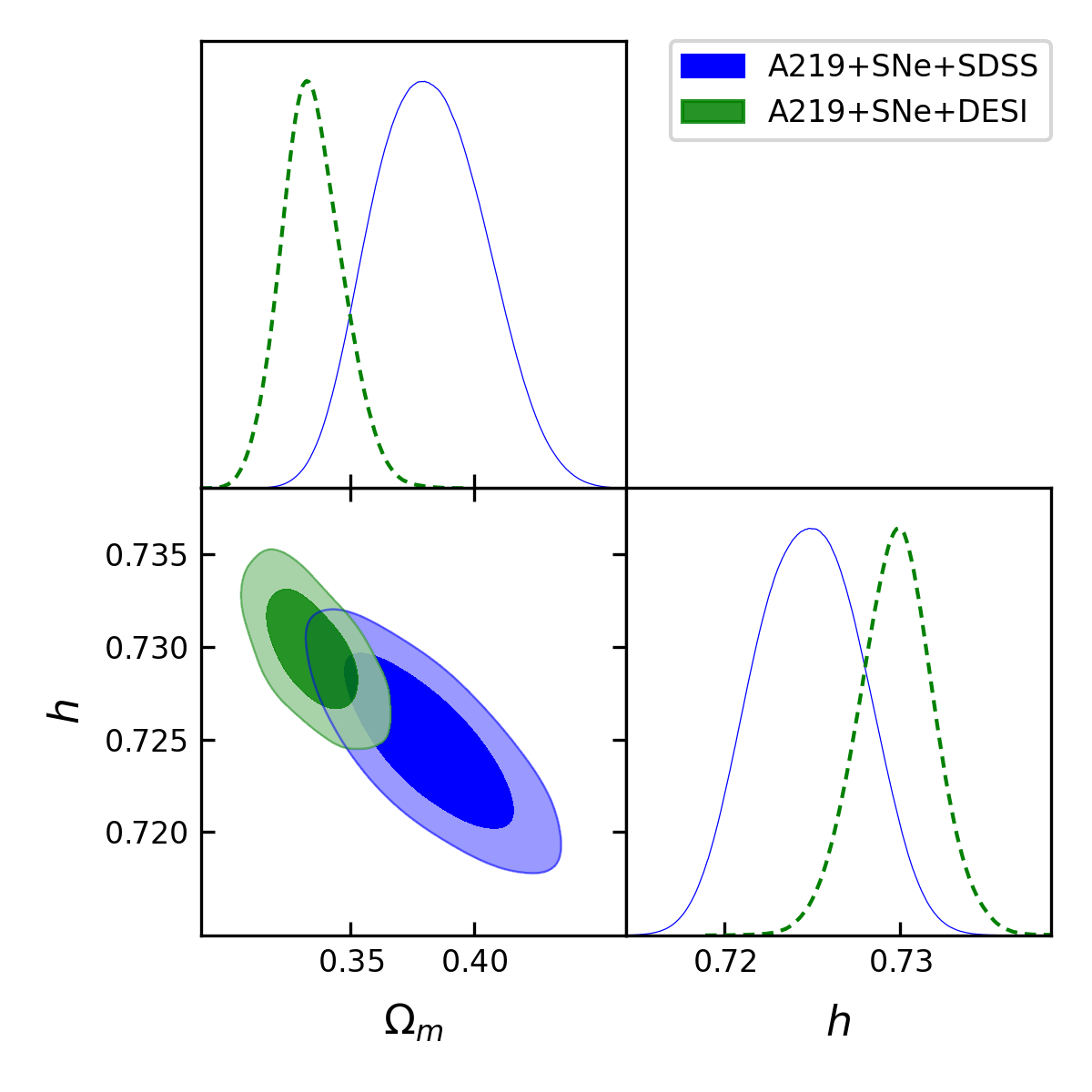}
	        \end{subfigure}
        \begin{subfigure}[b]{0.4\textwidth}
		        \centering
		        \includegraphics[width=\textwidth]{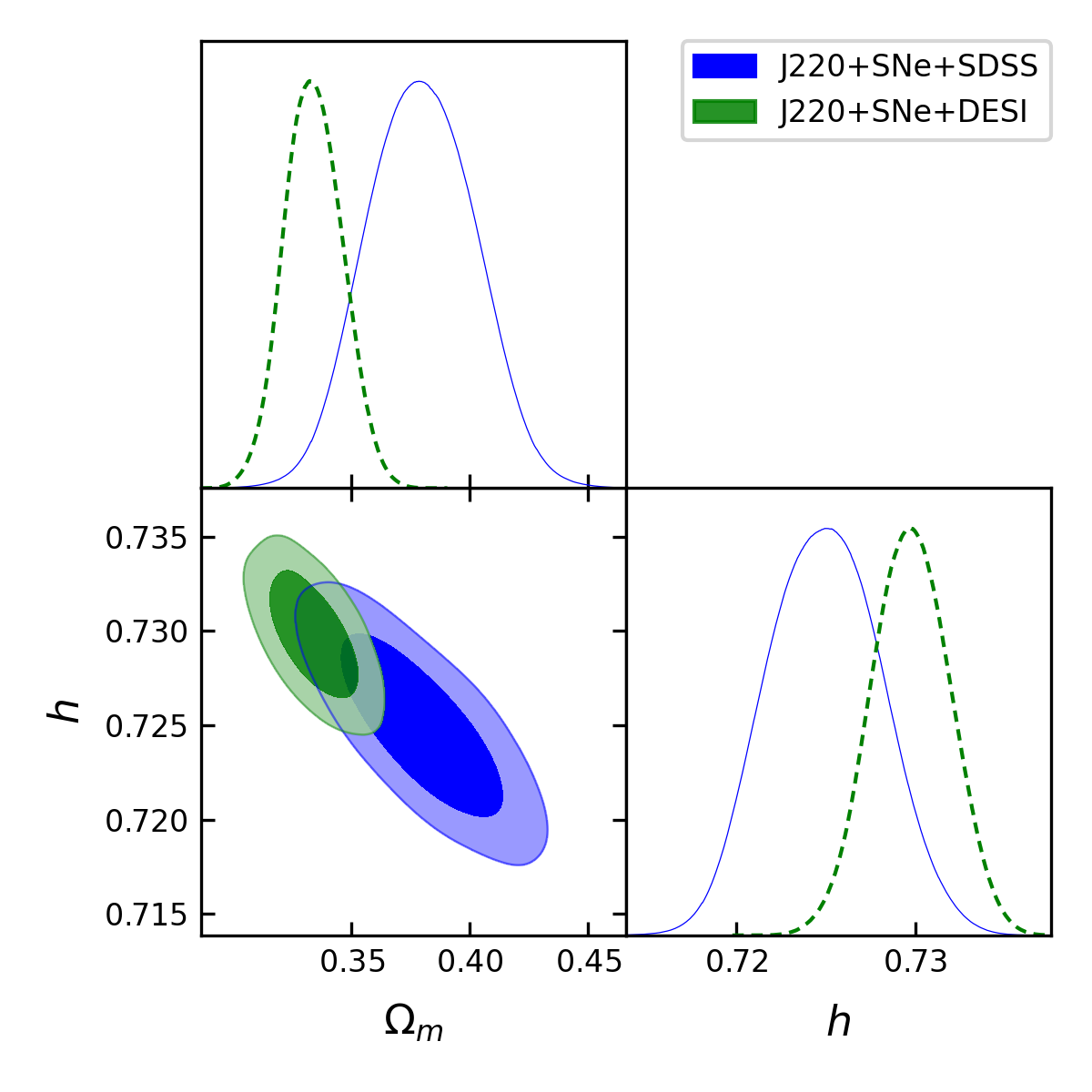}
	        \end{subfigure}	
		\caption{
			Joint constraints on parameters of $\Omega_m$, $h$ for the flat $\Lambda$CDM model combining   the A219 (\emph{left}) and J220 (\emph{right}) samples ($z>1.965$) with SNe and BAOs (SDSS and DESI).
		}
		\label{fig_GRB_SN_LCDM_N}
	\end{figure*}

\begin{figure*}
		\centering
		%		\includegraphics[width=0.35\linewidth]{./img/GRB_SN_BAO_wCDM.png}
%		\begin{subfigure}[b]{0.4\textwidth}
%	         \centering
%	         \includegraphics[width=\textwidth]{./img/constrain/wCDM_A219_SN_SDSS.png}
%        \end{subfigure}
%        \begin{subfigure}[b]{0.4\textwidth}
%         	\centering
%          	\includegraphics[width=\textwidth]{./img/constrain/wCDM_J220_SN_SDSS.png}
%        \end{subfigure}
%        \begin{subfigure}[b]{0.4\textwidth}
%	        \centering
%	        \includegraphics[width=\textwidth]{./img/constrain/wCDM_A219_SN_DESI.png}
%        \end{subfigure}
%        \begin{subfigure}[b]{0.4\textwidth}
%         	\centering
%	        \includegraphics[width=\textwidth]{./img/constrain/wCDM_J220_SN_DESI.png}
%        \end{subfigure}		
        \begin{subfigure}[b]{0.4\textwidth}
        	\centering
        	\includegraphics[width=\textwidth]{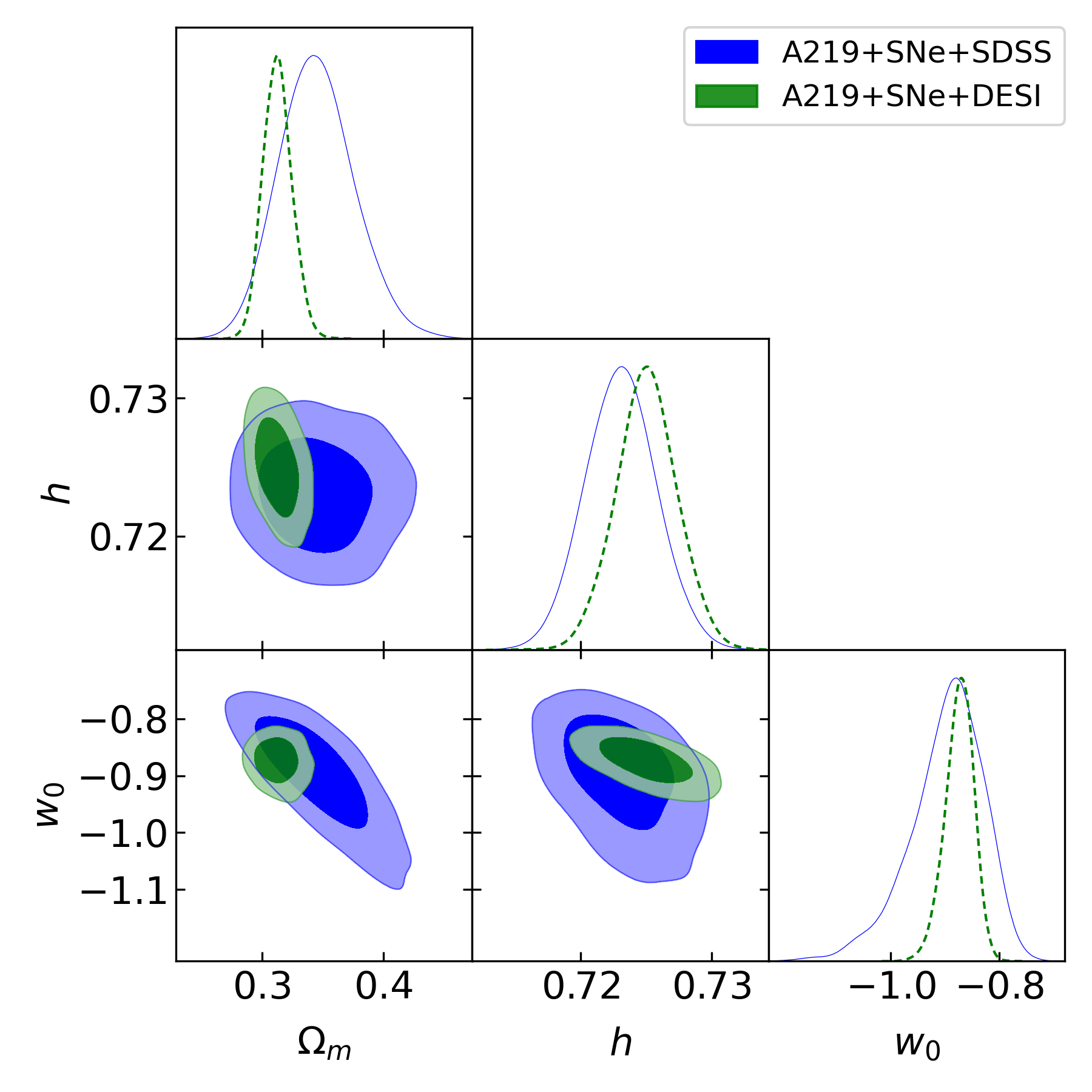}
        \end{subfigure}
        \begin{subfigure}[b]{0.4\textwidth}
        	\centering
        	\includegraphics[width=\textwidth]{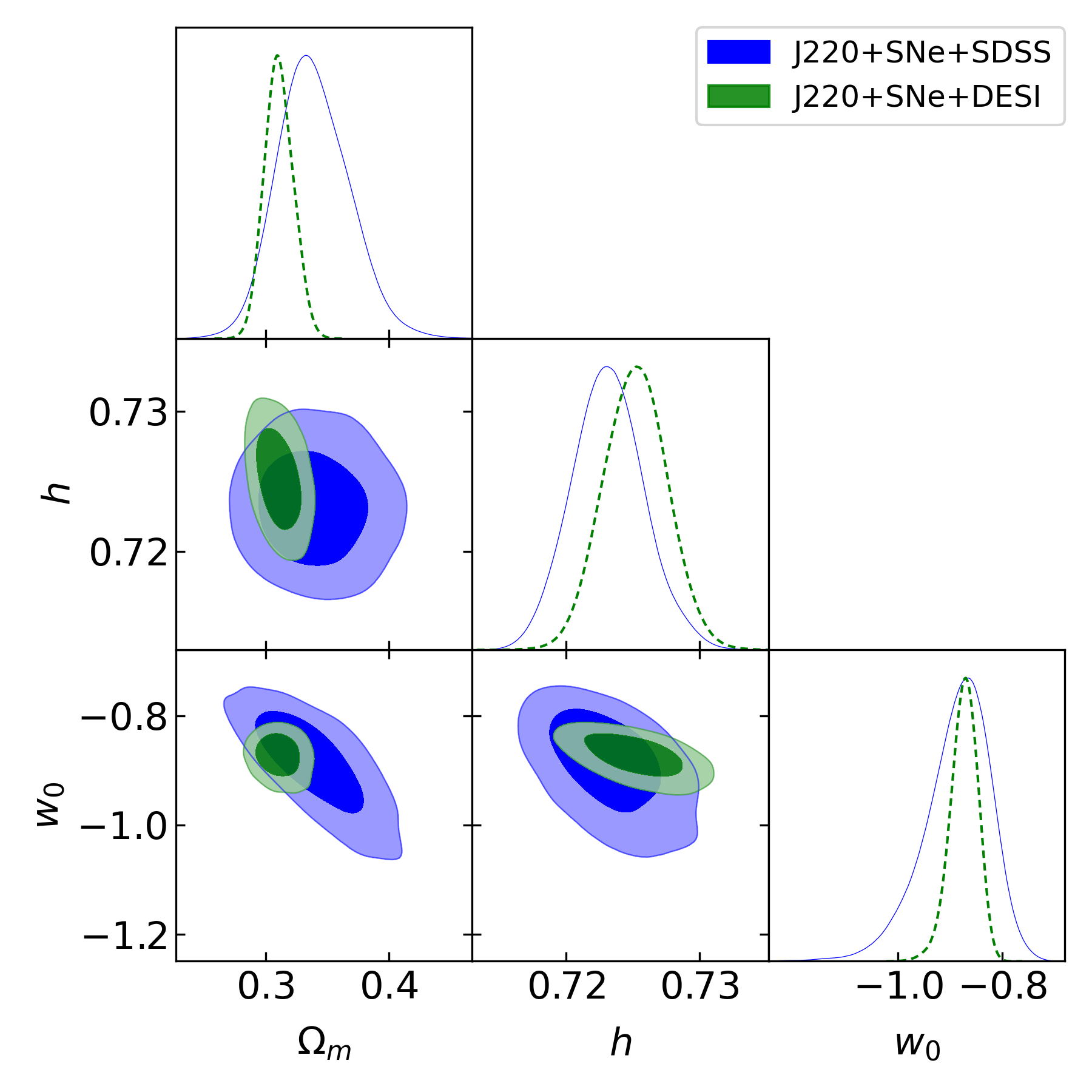}
        \end{subfigure}	
		\caption{Joint constraints on parameters of $\Omega_m$, $h$, and $w_0$  for the flat $w$CDM model combining  the A219 (\emph{left}) and J220 (\emph{right}) samples ($z>1.965$) with SNe and BAOs (SDSS and DESI).
		}
		\label{fig_GRB_SN_wCDM_N}
	\end{figure*}

\begin{figure*}
		\centering
		%		\includegraphics[width=0.35\linewidth]{./img/GRB_SN_BAO_CPL.png}
%		\begin{subfigure}[b]{0.4\textwidth}
%         	\centering
%         	\includegraphics[width=\textwidth]{./img/constrain/CPL_A219_SN_SDSS.png}
%        \end{subfigure}
%        \begin{subfigure}[b]{0.4\textwidth}
%	        \centering
%         	\includegraphics[width=\textwidth]{./img/constrain/CPL_J220_SN_SDSS.png}
%        \end{subfigure}
%        \begin{subfigure}[b]{0.4\textwidth}
%	       \centering
%	       \includegraphics[width=\textwidth]{./img/constrain/CPL_A219_SN_DESI.png}
%        \end{subfigure}
%        \begin{subfigure}[b]{0.4\textwidth}
%        	\centering
%        	\includegraphics[width=\textwidth]{./img/constrain/CPL_J220_SN_DESI.png}
%        \end{subfigure}	
       	\begin{subfigure}[b]{0.4\textwidth}
         	\centering
         	\includegraphics[width=\textwidth]{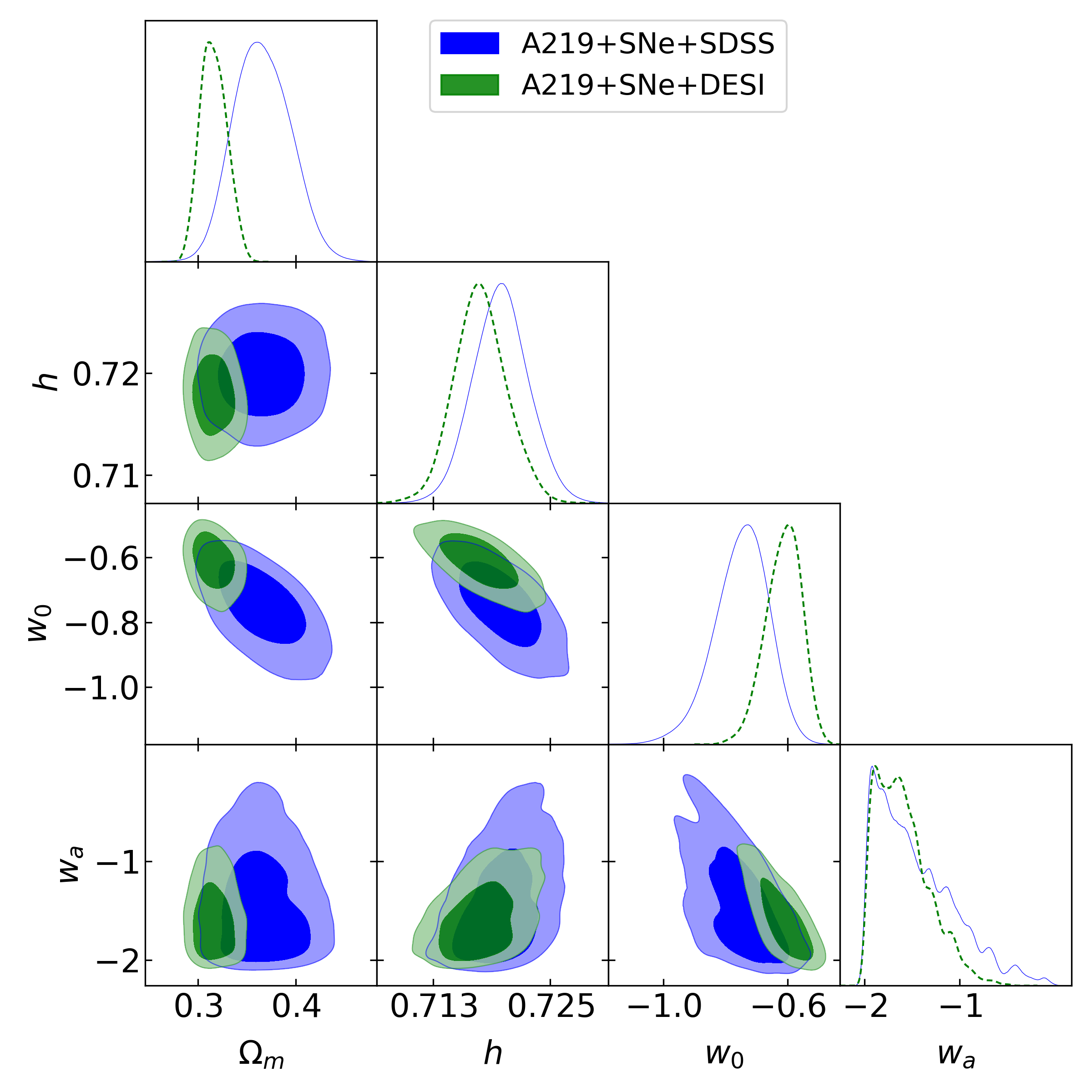}
        \end{subfigure}
        \begin{subfigure}[b]{0.4\textwidth}
         	\centering
         	\includegraphics[width=\textwidth]{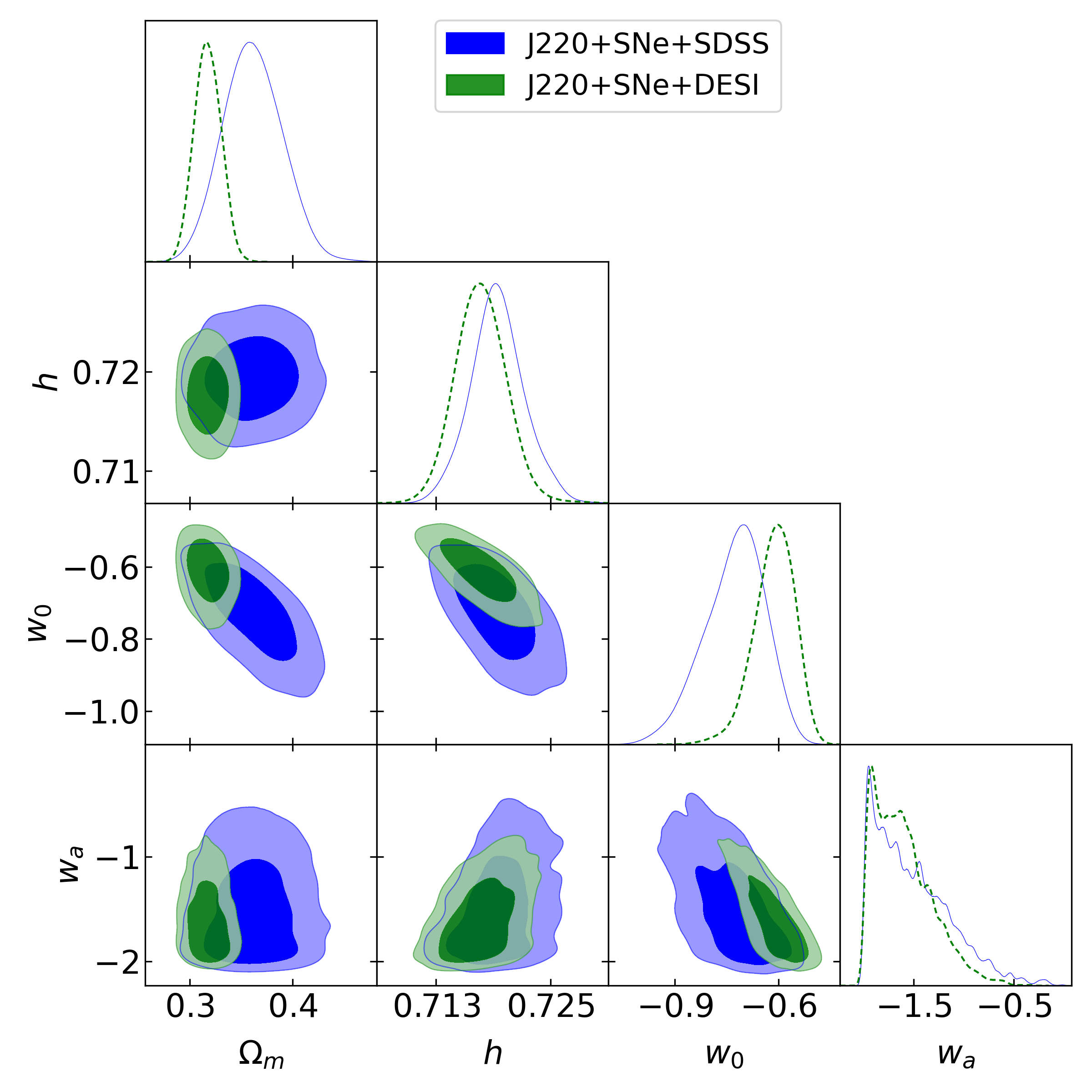}
        \end{subfigure}		
		\caption{Joint constraints on parameters of $\Omega_m$, $h$,  $w_0$ and $w_a$ for the flat CPL model combining the A219 (\emph{left}) and J220 (\emph{right}) samples ($z>1.965$) with SNe and BAOs (SDSS and DESI).
		}
		\label{fig_GRB_SN_CPL_N}
	\end{figure*}

Joint constraints combining GRBs with the A219  and J220 (\emph{right}) ($z\ge1.965$) and SNe, BAOs are shown in Fig. \ref{fig_GRB_SN_LCDM_N} ($\Lambda$CDM), Fig. \ref{fig_GRB_SN_wCDM_N} ($w$CDM) and Fig. \ref{fig_GRB_SN_CPL_N} (CPL), which are summarized in Table \ref{joint constraint results}.
Combining SNe Ia and BAOs with GRB samples, we find the results are consistent with the A219 and J220 samples.
For the $\Lambda$CDM model, we obtain $\Omega_{\rm m}=0.382^{+0.021}_{-0.024}$, $h=0.7248\pm{0.0030}$ and  $\Omega_{\rm m}=0.380\pm{0.022}$, $h=0.7250\pm{0.0031}$
with A219+SNe+SDSS and J220+SNe+SDSS at the 1$\sigma$ confidence level, respectively;
as well as $\Omega_{\rm m}=0.335^{+0.011}_{-0.013}$, $h=0.7298\pm{0.0022}$
and
$\Omega_{\rm m}=0.334\pm{0.012}$, $h=0.7298\pm{0.0022}$  with A219+SNe+DESI and J220+SNe+DESI at the 1$\sigma$ confidence level, respectively.
We %obtain %$\Omega_{\rm m}$ = $0.369^{+0.029}_{-0.029}$, $h$ = $0.6849^{+0.0028}_{-0.0028}$,
%$w_0$ = $-0.74^{+0.11}_{-0.07}$, $w_a$ = $-1.43^{+0.19}_{-0.54}$  with the A219 sample, and %$\Omega_{\rm m}$ = $0.376^{+0.027}_{-0.031}$, $h$ = $0.6849^{+0.0029}_{-0.0029}$,
%$w_0$ = $-0.76^{+0.11}_{-0.08}$, $w_a$ = $-1.39^{+0.24}_{-0.58}$  with the J220 sample at the 1$\sigma$ confidence level, respectively; %The results indicate that the $\Lambda$CDM model is preferred over the $w$CDM and CPL models.
find there is evident that %the $\Lambda$CDM model remains the best fit to the observational data, while
the results of the $w$CDM model $w_0\neq-1$ %($w_0$ = $-0.891^{+0.082}_{-0.055}$ with the A219 sample, and $w_0$ = $-0.905^{+0.092}_{-0.055}$ with the J220 sample)
and the CPL model  %($w_0$ = $-0.744^{+0.11}_{-0.068}$, $w_a$ = $-1.43^{+0.19}_{-0.54}$  with the A219 sample, and $w_0$ = $-0.757^{+0.11}_{-0.079}$, $w_a$ = $-1.39^{+0.24}_{-0.58}$  with the J220 sample) %suggest that although data
somewhat support  EoS ($w_0>-1$) variations with redshift ($w_a\neq0$).
%	, while $w_0<-1$ with GRBs+SNe+DESI, respectively.}
We also find the constraint results with the GRB samples at $z>1.965$ are consistent with the GRB samples at $z>1.4$.

$H_0$ with a redshift evolving is an interesting idea \citep{Dainotti2021,Dainotti2025} for the well-known $H_0$ tension \citep{Hu2023}. We compare to the fitting results from  CMB data based on the $\Lambda$CDM model at  very high-redshift ($H_0$ = 67.36 km $s^{-1}$ ${{\rm Mpc}}^{-1}$, $\Omega_m$ = 0.315)\citep{Plank2020} and SNe Ia at very low-redshift ($H_0$ = 74.3 km $s^{-1}$ ${{\rm Mpc}}^{-1}$, $\Omega_m$ = 0.298)\citep{Scolnic2022}. % to find that the $H_0$  value with GRBs alone at $z\ge1.965$ seems to favor the one from the Planck observations;
Because of the degeneracy between $M_B$ and $H_0$, a high $M_B$ inherently leads to a higher $H_0$. For example,
$\Delta M_B\approx0.10$ correspond to $\Delta H_0\approx3.0$.
Adding BAO at relative low redshift  to joint constraints
will also contributes to a higher $H_0$.
%The $\Omega_{\rm m}$ value of our results for the flat $\Lambda$CDM model is consistent with the CMB observations at the 1$\sigma$ confidence level.
We also find $\Omega_{\rm m}\approx0.40$ for the $\Lambda$CDM model with GRBs at relative high redshift, %which is consistent with the CMB observations at the 1$\sigma$ confidence level;
while adding BAO at relative low redshift will remove this trend \citep{Wang2024}.
\begin{table*}
	\centering
	\setlength{\tabcolsep}{2pt} % 减小列间距
	\renewcommand{\arraystretch}{1.4} % 调整行间距
	\caption{
		Joint constraints on parameters of $\Omega_m$, $h$, $w_0$ and $w_a$ for the flat $\Lambda$CDM model, the flat $w$CDM model, and the flat CPL model combining GRBs with the A219  and J220 samples ($z>1.965$) with SNe, and BAOs (SDSS and DESI).
	}
	\label{joint constraint results}
	\begin{tabular}{cccccccccc}
		\toprule
		Models & Data Set & $\Omega_{m}$ & $h$ & $w_0$ & $w_a$ & $\Delta$AIC & $\Delta$BIC
		%			$-2\ln \mathcal{L_{\rm R}}$  & $\Delta \rm AIC$  & $\Delta \rm BIC$
		\\
		\hline
		\specialrule{0em}{1pt}{1pt}
		$\Lambda$CDM &
		A219+SNe+SDSS &
		$0.382^{+0.021}_{-0.024}$  &
		$0.7248\pm{0.0030}$  &
		- &
		- &
		0 &
		0
		\\
		 &
		A219+SNe+DESI &
		$0.335^{+0.011}_{-0.013}$  &
		$0.7298\pm{0.0022}$  &
		- &
		- &
		0 &
		0
		\\
        &
		J220+SNe+SDSS&
		$0.380^{+0.022}_{-0.022}$  &
		$0.7250\pm{0.0031}$  &
		- &
		- &
		0 &
		0
		\\
		 &
		J220+SNe+DESI &
		$0.334^{+0.012}_{-0.012}$  &
		$0.7298\pm{0.0022}$  &
		- &
		- &
		0 &
		0
		\\
		\hline
		\specialrule{0em}{1pt}{1pt}
		$w$CDM &
		A219+SNe+SDSS &
		$0.345^{+0.028}_{-0.033}$  &
		$0.7231\pm{0.0027}$ &
		$-0.90^{+0.08}_{-0.05}$ &
		- &
		2.85 &
		8.35
		\\
		&
		A219+SNe+DESI  &
		$0.313^{+0.011}_{-0.013}$  &
		$0.7250\pm{0.0023}$ &
		$-0.87^{+0.03}_{-0.02}$ &
		- &
		3.37 &
		8.87
		\\
        &
		J220+SNe+SDSS  &
		$0.339^{+0.026}_{-0.031}$  &
		$0.7232\pm{0.0027}$ &
		$-0.88^{+0.07}_{-0.05}$ &
		- &
		2.66 &
		8.16
		\\
		 &
		J220+SNe+DESI  &
		$0.311^{+0.012}_{-0.012}$  &
		$0.7252\pm{0.0024}$ &
		$-0.87^{+0.03}_{-0.02}$ &
		- &
		3.34 &
		8.86
		\\
		\hline
		\specialrule{0em}{1pt}{1pt}
		CPL                &
		A219+SNe+SDSS  &
		$0.366^{+0.027}_{-0.030}$  &
		$0.7200\pm{0.0029}$  &
		$-0.75^{+0.10}_{-0.07}$ &
		$-1.46^{+0.18}_{-0.52}$ &
		3.44 &
		14.45
		\\
		               &
		A219+SNe+DESI  &
        $0.316^{+0.013}_{-0.015}$  &
        $0.7179\pm{0.0027}$  &
        $-0.61^{+0.07}_{-0.05}$ &
        $-1.59^{+0.14}_{-0.36}$ &
        4.21 &
        15.22
        \\
         &
		J220+SNe+SDSS  &
		$0.361^{+0.029}_{-0.029}$  &
		$0.7194\pm{0.0029}$  &
		$-0.73^{+0.10}_{-0.07}$ &
		$-1.51^{+0.16}_{-0.48}$ &
		3.21 &
		14.20
        \\
        &
        J220+SNe+DESI  &
        $0.318^{+0.013}_{-0.013}$  &
        $0.7177\pm{0.0027}$  &
        $-0.61^{+0.07}_{-0.05}$ &
        $-1.60^{+0.13}_{-0.37}$ &
        4.01 &
        15.03
        \\
		\bottomrule
	\end{tabular}
\end{table*}

Finally, we compare the dark energy models using the Akaike Information Criterion (AIC) and Bayesian Information Criterion (BIC), which are defined as follows:$	\text{AIC}=-2\ln\mathcal{L}_{\rm max}+2k,~
\text{BIC}=-2\ln\mathcal{L}_{\rm max}+k\ln N $. where $\mathcal{L}_{\rm max}$ represents the maximum likelihood. For the Gaussian case, $\chi_{\rm min}^2=-2\ln\mathcal{L}_{\rm max}$. %The values of $\Delta\text{AIC}$ and $\Delta\text{BIC}$ are given by: $	\Delta\text{AIC}=\Delta\chi_{\rm min}^2+2\Delta k,~	\Delta\text{BIC}=\Delta\chi_{\rm min}^2+\Delta k\ln N.$
The results for the values of $\Delta\text{AIC}$ and $\Delta\text{BIC}$ relative to the reference model (the $\rm \Lambda$CDM model) are summarized in Table \ref{joint constraint results}. We find that the $\rm \Lambda$CDM model is preferred over the $w$CDM and CPL models, which is consistent with previous analyses \citep{Amati2019,Montiel2021,LZL2023,Wang2024,WL2024} and \citep{Cao2024,Jia2022}.

	\section{Conclusions}		
	Throughout this work, we utilize the latest OHD data to reconstruct $H(z)$ through ANN to calibrate the Amati relation at low redshift with the A219  and J220 samples to construct the GRB Hubble diagram.
We consider the physical relationships between the data to introduce the covariance matrix and KL divergence of the data.
We find the results are consistent with the A219 and J220 samples combining SNe Ia and BAOs (SDSS or DESI). %; as well as  combining SDSS and DESI with GRB+SNe Ia.
% we find the results are consistent with the A219 and J220 samples.
For the $\Lambda$CDM model, we obtain $\Omega_{\rm m}=0.382^{+0.021}_{-0.024}$, $h=0.7248\pm{0.0030}$ and
$\Omega_{\rm m}=0.335^{+0.011}_{-0.013}$, $h=0.7298\pm{0.0022}$
with A219+SNe+SDSS and A219+SNe+DESI;
as well as  $\Omega_{\rm m}=0.380\pm{0.022}$, $h=0.7250\pm{0.0031}$ and
$\Omega_{\rm m}=0.334\pm{0.012}$, $h=0.7298\pm{0.0022}$  with J220+SNe+SDSS and J220+SNe+DESI at the 1$\sigma$ confidence level, respectively.
We %obtain %$\Omega_{\rm m}$ = $0.369^{+0.029}_{-0.029}$, $h$ = $0.6849^{+0.0028}_{-0.0028}$,
%$w_0$ = $-0.74^{+0.11}_{-0.07}$, $w_a$ = $-1.43^{+0.19}_{-0.54}$  with the A219 sample, and %$\Omega_{\rm m}$ = $0.376^{+0.027}_{-0.031}$, $h$ = $0.6849^{+0.0029}_{-0.0029}$,
%$w_0$ = $-0.76^{+0.11}_{-0.08}$, $w_a$ = $-1.39^{+0.24}_{-0.58}$  with the J220 sample at the 1$\sigma$ confidence level, respectively; %The results indicate that the $\Lambda$CDM model is preferred over the $w$CDM and CPL models.
find there is evident that %the $\Lambda$CDM model remains the best fit to the observational data, while
the results of the $w$CDM and CPL models %suggest that although data
somewhat support  EoS ($w_0>-1$) variations with redshift ($w_a\neq0$). %\textbf{and $w_0>-1$ with the joint constraints.}
%	, while $w_0<-1$ with GRBs+SNe+DESI, respectively.}
%	Combining GRB data at $z \ge 1.4$ with  SNe Ia and BAOs by the MCMC method, we obtain
%	$\Omega_{\rm m}$ = $0.369^{+0.029}_{-0.029}$, $h$ = $0.6849^{+0.0028}_{-0.0028}$,  $w_0$ = $-0.744^{+0.11}_{-0.068}$, $w_a$ = $-1.43^{+0.19}_{-0.54}$  with the A219 sample, and $\Omega_{\rm m}$ = $0.376^{+0.027}_{-0.031}$, $h$ = $0.6849^{+0.0029}_{-0.0029}$,  $w_0$ = $-0.757^{+0.11}_{-0.079}$, $w_a$ = $-1.39^{+0.24}_{-0.58}$  with the J220 sample at the 1$\sigma$ confidence level, respectively;
%which indicates that the current GRB, SNe, and BAO datasets provide precise constraints on DE models. %, but there is still some uncertainty in the parameter space.
%\textbf{A high $M_B$ inherently leads to a higher $H_0$ because of the degeneracy between them.}
Adding BAO at relative low redshift  to joint constraints will contributes to a higher $H_0$. However, the $\Lambda$CDM model is preferred over the $w$CDM and CPL models using the AIC and BIC.
	
	More recently, \cite{WL2024} presented a sample of long GRBs  from 15 years of the Fermi-GBM catalogue with identified redshift, in which the GOLD sample contains 123 long GRBs at $z\le5.6$ and the FULL sample contains 151 long
	GRBs with redshifts at $z\le8.2$.
	\cite{Jiang2024} find new features by replacing the Pantheon+ with the Union3 \citep{Rubin2023} and the full five years of the Dark Energy Survey (DES) Supernova Program \citep{Abbott2024} to reconstruct the late-time expansion history in light of the latest BAO measurements from DESI \citep{DESI2024} based on Gaussian processes.
	\cite{Alfano2024a,Alfano2024b}  consider DESI 2024 data to constrain dark energy models with the Amati relation, Yonetoku relation \citep{Yonetoku2004} and the Combo relation \citep{Izzo2015} by the B\'{e}zier parametric approach;
as well as the Dainotti relations \citep{Dainotti2008,Dainotti2016,Dainotti2020} as standard candles can also be calibrated  model-independently	\citep{Favale2024,Alfano2024c,Sethi2024,Zhang2025,Mukherjee2024b}.
In the following work, we expect that adding the 15 years of the Fermi-GBM catalogue at $0.0785\le z\le8.2$, %\citep{WL2024},
the SNe Ia from the Union3 sample with 2087 SNe Ia at $0.01 < z < 2.26$ %\citep{Rubin2023}
and the 5-year DES with 1635 SNe Ia at $0.1 < z < 1.3$, %\citep{Abbott2024},
as well as  BAO measurements from DESI data release %\citep{DESI2024}
will lead to more accurate constraints on the cosmological parameters and the evolutionary behavior of DE.

	%	(Comparative constraints on cosmological parameters and model comparison)In this work, The Amati relation are calibrated from the latest OHD with the CC method \citep{Jiao2022} using machine learning at $z<1.4$ to obtain GRBs at $z\ge1.4$ which can be used to constrain cosmological models.
	%	By the MCMC method  with ? and ?, we obtain  $\Omega_{\rm m} = $ and $H_0 = $ for the ?model:  $\Omega_{\rm m} = $ and $H_0 = $ for ?model.
	%	We also find that $\Lambda$CDM model is preferred with respect to the $w$CDM and the CPL models.	
	%	(Compare with the results obtained by using the Gaussian process before.)Compare with the previous works which use GP method to calibrate GRBs\cite{Liang2022,LZL2023,Xie2023,Zhang2023,Wang2024}, we get .....	
	%	(machine learning part result, Highlight the results and conclusions of machine learning, improve and enhance. )... However, the result we get shows an consistency with \cite{WangGJ2020}.
	%	(conclusion)

	\section*{Acknowledgments}
We thank Yang Liu, Prof. Puxun Wu, Prof.  Jianchao Feng  for kind help and discussions. We also thank the anonymous referee for  helpful comments and constructive suggestions.	This project was supported by the Guizhou Provincial Science and Technology Foundation: QKHJC-ZK[2021] Key 020 and QKHJC-ZK[2024] general 443. Y. Liu was supported by the NSFC under Grant No. 12373063.
	
	%\section*{DATA AVAILABILITY}
	%Data are available at the following references: 33 OHD obtained with the CC method from Table \ref{OHD} and references therein, the J221 sample of GRB data set from \cite{Jia2022} and the Pantheon+ sample from \cite{Scolnic2022}. %The data underlying this article will be shared on reasonable request to the corresponding author.
	
	%\bibliographystyle{plain}
	%
	%\bibliography{refs}
	
\end{document}